\documentclass[prd,tightenlines,floatfix,showpacs,preprintnumbers,nofootinbib,eqsecnum]{revtex4}
  \usepackage{amssymb}
   \usepackage{amsmath} 
    \usepackage{amsfonts}
     \usepackage{epsfig}
     \usepackage{color}
      \usepackage{bm} 

\usepackage{booktabs}
\usepackage{multirow}
\usepackage{slashed}
\usepackage{comment}

\usepackage{cancel}
\usepackage[normalem]{ulem}

\usepackage{amscd}
\usepackage{graphicx}
\usepackage{subfigure}
\usepackage{hyperref}

\def\doublespace{\def\baselinestretch{1.6}\large\normalsize}
\def\normalspace{\def\baselinestretch{1.0}\normalsize}

\def\Caption#1{
  \normalspace
  \vskip-1mm\caption{\sl#1}\vskip-1mm
  \doublespace
}

\def\BA{\begin{eqnarray}}
\def\BE{\begin{equation}}
\def\BF{\begin{figure}[htb]}
\def\BT{\begin{table}[htb]}
\def\EA{\end{eqnarray}}
\def\EE{\end{equation}}
\def\EF{\end{figure}}
\def\ET{\end{table}}

\def\ra{\rangle}
\def\fm{\,\mbox{fm}}

\def\GeV{\,\mbox{GeV}}

\def\Jpsi{J\!/\!\psi}
\def\psip{\psi^{\,\prime}}

\def\sqq{\sigma_{Q\bar Q}}

\def\aem{\alpha_{em}}

\def\lsim{\mathrel{\rlap{\lower4pt\hbox{\hskip1pt$\sim$}}
     \raise1pt\hbox{$<$}}}         
 \def\gsim{\mathrel{\rlap{\lower4pt\hbox{\hskip1pt$\sim$}}
     \raise1pt\hbox{$>$}}}         

\begin{document}

\title{
Non-linear effects in
coherent photoproduction of heavy quarkonia off nuclei: 
\\
Green's function
formalism vs Balitsky-Kovchegov equation
\vspace*{0.20cm}
}

\author{I. Koval$^{1}$}
\email{kovalill@fjfi.cvut.cz}

\author{J. Nemchik$^{1,2}$}
\email{jan.nemcik@fjfi.cvut.cz}

\author{J. \'Obertov\'a$^{1}$}
\email{jaroslava.obertova@fjfi.cvut.cz}

\affiliation{
\vspace*{0.20cm}
$^1$
Czech Technical University in Prague, FNSPE, B\v rehov\'a 7, 11519
Prague, Czech Republic}
\affiliation{$^2$
Institute of Experimental Physics SAS, Watsonova 47, 04001 Ko\v sice, Slovakia
}

\date{\today}
%
%
\begin{abstract}
%
%
\vspace*{5mm}
We present an analysis of QCD non-linear (NL) effects 
in the coherent photoproduction of charmonia off nuclei 
at high energies corresponding to future measurements at the LHC. 
For the first time,
we confront the energy evolution of such effects
within our rigorous Green's function (GF) formalism with results 
from Refs.~\cite{Cepila:2025ujv,Cepila:2024qge}, which are
based on the solution of the leading-order (LO) and next-to-leading-order
(NLO) Balitsky-Kovchegov (BK) evolution equations.
We demonstrate a huge difference in the NL effects between these
two formalisms, exhibiting a rather strong rise with photon energy.
Our analysis shows that NLO calculations of the BK equation 
from Ref.~\cite{Cepila:2024qge}
slow down the rate of the high-energy LO evolution 
of NL effects.
Although these calculations reduce the differences between the GF and BK predictions, 
they still lead to a much stronger onset of leading-twist shadowing effects 
compared to the GF formalism.
The BK equation in its present form cannot be
applied to nuclear targets and thus requires a non-trivial modification 
that properly incorporates quantum coherence effects. 

\end{abstract}
%
%


\maketitle

%
%
%
\section{Introduction}
\label{Sec:intro}
%
%
%

Gluon shadowing represents an integral part of nuclear shadowing in interactions 
with nuclear targets at high energies (see review 
\cite{Paakkinen:2026dnh} and references therein).
A deeper understanding of the role of quantum coherence in 
nuclear shadowing effects is crucial for a proper interpretation of empirical
results from photon-nucleus, hadron-nucleus and heavy-ion experiments
\cite{Kopeliovich:1991pu,Kopeliovich:1998gv,Kopeliovich:1999am,Kopeliovich:2000ra,Kopeliovich:2001ee,Kopeliovich:2001xj,Ivanov:2002kc,Nemchik:2003wx,Kopeliovich:2008ek,Krelina:2020ipn,Kopeliovich:2022jwe,Kopeliovich:2020has}.
%
The nuclear shadowing is typically studied in deep-inelastic scattering (DIS) off nuclei 
in terms of the nuclear structure functions. 
At large photon energies, corresponding to small Bjorken $x < 0.01$, the structure function $F_2$ per nucleon is found to be smaller in nuclei than in a free nucleon 
\cite{Arneodo:1992wf}.

The interpretation of nuclear shadowing is reference-frame dependent. 
In the infinite-momentum frame of the nucleus,
it can be viewed as a result of parton fusion 
\cite{Kancheli:1973vc,Gribov:1983ivg,Mueller:1985wy,Qiu:1986wh}. 
This leads to a reduction of the parton density at low $x$. 
In the rest frame of the nucleus, however, this phenomenon can be 
treated as nuclear shadowing of the hadronic fluctuations of the virtual photon, 
generated by their multiple scatterings inside the target 
\cite{Bauer:1977iq,Frankfurt:1988nt,Brodsky:1989qz,Brodsky:2004qa,Nikolaev:1990ja,Melnitchouk:1993vc,Nikolaev:1994de,Piller:1995kh,Kopeliovich:1995yr,Kopeliovich:1995ju,Piller:1999wx,Kopeliovich:2000ra,Nemchik:2003wx,Kopeliovich:2008ek,Krelina:2020ipn}.
%
Although both interpretations complement each other, the latter one is more intuitive and straightforward.

Recent experiments involving ultra-peripheral collisions (UPCs) at the Large Hadron 
Collider (LHC) represent a highly effective tool for investigating nuclear shadowing.
Specifically, the exclusive photoproduction of heavy quarkonia off nuclei provides a 
powerful means to probe the gluon shadowing phenomenon arising from the multi-gluon 
components of the incoming, quasi-real photon.
Such components represent the dominant source of nuclear shadowing 
\cite{Kopeliovich:2022jwe}.
%
Moreover, gluon shadowing 
effects will be further investigated by future experiments at the planned Large 
Hadron-Electron Collider (LHeC) operating at the High Luminosity-LHC 
\cite{LHeC:2020van} 
at even higher center-of-mass (c.m.) energies and with virtual photons. 
Since gluon shadowing corresponds to a non-linear term in the evolution 
equations, the coherent and incoherent electroproduction of heavy quarkonia off nuclei 
enables a systematic study of the significance of gluon saturation effects 
(see
Ref.~\cite{Nemchik:2025myg}
and references therein). 
Finally, the study of gluon 
shadowing can help rule out various phenomenological models, while simultaneously 
addressing the applicability of standard evolution equations to processes on 
nuclear targets.

The dynamics of nuclear shadowing is governed by quantum coherence effects, 
which result from the destructive interference of amplitudes 
for which 
the interaction takes place on different bound nucleons. 
The wave function of the photon can be expanded over various Fock states 
containing either bare quarks or states with additional numbers of gluons. 
The quantum coherence can be physically interpreted as the lifetime of a particular photon fluctuation. 
Each Fock component contributes separately to the total nuclear shadowing 
in accordance with its corresponding coherence length. 
Consequently, quark and gluon shadowing are associated with
the lowest $|Q\bar Q\ra$ Fock state 
and with the higher photon components containing gluons, respectively.

Treating photonuclear reactions ($Q^2 \approx 0$) and considering symmetric 
lowest $Q\bar Q$ photon fluctuations, where the quarks carry approximately equal 
fractions of the photon momentum, the corresponding fluctuation lifetime 
can be quantified in terms of the coherence length (CL), which takes the form, 
%
\BE 
l_c^{Q\bar Q}
= 
\frac{W^2-m_N^2}{m_N M_{Q\bar Q}^2}
\approx
\frac{W^2-m_N^2}{m_N M_V^2}
\,,
  \label{lc-qq}
\EE
%
where $W$ is the c.m. energy of the photon-nucleon system,
$M_{Q\bar Q}$ is the effective mass of the $Q\bar Q$ pair, and
$m_N$ and $M_V$ denote the masses of the nucleon and vector meson, respectively. 
The onset 
of quark shadowing requires a sufficiently long coherence length, such that 
$l_c^{Q\bar{Q}} \gsim R_A$, where $R_A$ is the nuclear radius.

In the diffractive electroproduction of heavy quarkonia, the quark shadowing 
is substantially suppressed due to large quark mass and thus acts as 
a higher-twist effect
\cite{Kopeliovich:2022jwe}.
%
In contrast, the gluon shadowing 
effect is of leading-twist origin because the transverse 
size of the multi-gluon Fock states is relatively large, nearly scale independent, 
and governed by the gluon propagation radius, $r_G \sim 0.3\,\fm$ 
\cite{Kopeliovich:1999am}.
%
Consequently, gluon shadowing is much stronger than higher-twist shadowing, 
which is controlled by the small size of the $Q\bar{Q}$ dipoles.
The corresponding coherence (radiation) length for 
the one-gluon Fock state $|Q\bar QG\ra$ of the photon
takes the form,
%
\BE
l_{c}^G
=
\frac{W^2-m_N^2}{m_N M_{Q\bar QG}^2}
=
\frac{(W^2 - m_N^2)\cdot\alpha_G(1-\alpha_G)}
{m_N\Bigl [k_T^2+(1-\alpha_G)\,m_G^2+\alpha_G M_{Q\bar Q}^2 \Bigr] }\,, 
  \label{lg-full}
\EE
%
where $k_T$ is the gluon transverse momentum,
$\alpha_G$ is the light-front (LF) fraction of the photon momentum carried by the gluon,
$M_{Q\bar QG}$ is the effective mass of the ${Q\bar QG}$ photon fluctuation, 
and 
the effective gluon mass  $m_G\sim 1/r_G \approx 0.65\text{--}0.70\GeV$ is fixed by data on gluon radiation 
\cite{Kopeliovich:1999am,Kopeliovich:2007pq}. 
%
Similar to quark shadowing, the onset of gluon shadowing
requires a sufficiently long $l_c^G\gsim d$, 
where $d\approx 2\,\fm$ is the mean separation between bound nucleons. 
One can also see that $l_c^G\ll l_c^{Q\bar Q}$ due to the strong inequality
$M_{Q\bar Q}\ll M_{Q\bar QG}$ and the small momentum fraction $\alpha_G\ll 1$, in 
agreement with calculations in
Ref.~\cite{Kopeliovich:2000ra}.
%
The contribution of gluon suppression to nuclear shadowing becomes effective 
at small $x < 0.01$ 
\cite{Kopeliovich:1999am}.
The corresponding energy range is safely covered by recent UPC experiments 
at the LHC and will be further accessed by future measurements at the LHeC.

Higher multi-gluon Fock components of the photon possess even shorter lifetimes and
preserve the hierarchy $l_c^{nG}\ll l_c^{(n-1)G}\ll \dots \ll l_c^{2G}\ll l_c^G$, 
since the corresponding effective masses satisfy the strong inequalities 
$M_{Q\bar QnG}\gg M_{Q\bar Q(n-1)G}\gg \dots \gg M_{Q\bar Q2G}\gg M_{Q\bar QG}$. 
Therefore, as more gluons are included in the photon fluctuation, 
higher photon energies are required for the shadowing 
effects related to a particular Fock state to manifest themselves 
(see Sec. IV.B in
Ref.~\cite{Kopeliovich:2022jwe}).

Different (but equivalent) descriptions of gluon shadowing (GS) can be formulated 
depending on the reference frame. 
In the rest frame of the nucleus, this phenomenon appears as a specific part 
of Gribov's inelastic corrections 
\cite{Gribov:1968jf}.
%
Conversely,
in the infinite-momentum frame of the nucleus, GS can be viewed as the fusion 
of gluons that overlap in the longitudinal direction at small $x$, 
leading to a subsequent reduction in the gluon density.

Due to the current lack of direct experimental data on gluon shadowing, one must 
rely on theoretical model calculations (see compilation in
Ref.~\cite{Paakkinen:2026dnh}).
%
In most available models, the final results for nuclear shadowing have been obtained 
only within the high-energy eikonal (``frozen'') approximation, where 
$l_c^{Q\bar Q}, l_c^{G}\gg R_A$, 
or eventually by incorporating  finite coherence length corrections 
through a static part of the CL included in the corresponding phase factor.
However, when considering higher 
multi-gluon Fock states of the photon,  
the corresponding $Q\bar Q\text{--}G$, $\dots$ dipoles 
cannot be treated as ``frozen'' even at very high 
energies due to a divergent $d\alpha_G/\alpha_G$ behavior.
To overcome this limitation, more realistic calculations 
of GS must be performed beyond the ``frozen'' approximation by utilizing the 
path-integral technique
\cite{Kopeliovich:1999am,Ivanov:2002kc,Kopeliovich:2008ek,Krelina:2020ipn,Kopeliovich:2022jwe}.
%
In this framework, the coherence length is dynamically 
regulated by a varying dipole effective mass within the 
evolution equation for the Green's function.
This approach provides a rigorous treatment 
of quantum coherence effects compared to alternative models and exerts a direct 
impact on the predicted magnitude of gluon shadowing.

The corresponding results of the GS contribution to nuclear suppression 
have been obtained within the light-front 
quantum chomodynamics (QCD) dipole approach. 
It was found that GS represents rather strong effect in both coherent and 
incoherent production of vector mesons
\cite{Kopeliovich:2001xj,Nemchik:2002ug,Ivanov:2002kc,Kopeliovich:2020has,Kopeliovich:2022jwe,Nemchik:2024lny,Nemchik:2025myg},
in DIS 
\cite{Kopeliovich:2000ra,Nemchik:2003wx,Kopeliovich:2008ek,Krelina:2020ipn},
and also in production of Drell-Yan pairs 
\cite{Kopeliovich:2001hf}.

The reduction in gluon density induced by GS effects corresponds to a 
non-linear term in the QCD evolution equations
\cite{Gribov:1981ac,Gribov:1983ivg,balitsky,kovchegov}.
%
This provides a strong motivation to investigate the energy behavior of shadowing effects 
by confronting our calculations, which are based on a rigorous Green's function 
formalism, with results obtained from explicit solutions of the evolution equations.
Recently, new predictions for the onset of NL effects in the coherent photoproduction 
of charmonia off nuclei were presented in 
Ref.~\cite{Cepila:2025ujv}.
%
In that work,
within the light-front color dipole framework, the QCD energy evolution of the dipole amplitude was obtained by solving the LO 
Balitsky-Kovchegov evolution 
\cite{balitsky,kovchegov}
with the running coupling (rc) and collinearly improved (CI) Kernel.
Furthermore, a very recent work
\cite{Cepila:2024qge}
reported NLO calculations of the BK equation, thus providing
valuable new insights into the energy dependence of these NL effects.
This offers an 
additional incentive to compare those findings with our predictions based on the 
path-integral technique.
Such a comparative analysis of two fundamentally different 
approaches to energy evolution of NL effects is of paramount importance for evaluating the significance of gluon saturation effects. 
To the best of our knowledge, except for the results from
Ref.~\cite{Cepila:2024qge},
%
no other LO and NLO calculations that simultaneously account 
for and neglect NL effects are currently available for a direct comparison with our 
GF framework.

In this article, for simplicity, we examine the role of leading-twist corrections with 
respect to the coherent photoproduction of charmonia on a lead target.
Such a reaction 
allows us to minimize theoretical uncertainties and test our predictions given the 
abundance of available experimental data.
Other recent NLO calculations of exclusive heavy quarkonium production 
(see e.g., 
Ref.~\cite{Mantysaari:2022kdm})
have been performed solely for a proton target and thus are not yet applicable 
for the study of QCD phenomena accompanying photoproduction processes on nuclear targets.
Furthermore, the NLO calculations of charmonium production in UPCs 
\cite{Eskola:2022vaf,Eskola:2022vpi} 
do not accurately incorporate the quantum coherence effects across the entire rapidity region.
In those studies, NL effects were included using nuclear parton distribution 
functions from the recent global data fit, which provides only the gluon shadowing 
integrated over the nuclear impact parameter. 
In a recent review paper
\cite{Paakkinen:2026dnh}
the leading-twist approximation for nuclear shadowing within the Gribov-Glauber theory
leads to significantly smaller values for the nucleus-to-nucleon ratios of the 
proton structure functions
(i.e. to much stronger shadowing effects) compared to our results in
\cite{Krelina:2020ipn}
obtained within the QCD dipole approach
based on the Green's function formalism.
Such discrepancies in predictions from various models can 
be effectively tested by studying heavy quarkonium photoproduction off nuclei at high 
energies, as performed in the present paper, where higher-twist effects are 
substantially reduced.

The paper is organized as follows.
In Sec.~\ref{gf}, we present a brief review of the light-front 
dipole formalism along with the corresponding formula for the charmonium cross section 
in coherent photoproduction off nuclei, $\gamma A \to J/\psi A$.
Here, we also present 
a rigorous path-integral formalism for the lowest $Q\bar{Q}$ Fock state of the photon, 
demonstrating that this framework naturally leads to the standard eikonal expression for 
the coherent production cross section in the high-energy limit. 
Since leading-twist 
shadowing corrections represent a dominant effect at high photon energies, in Sec.~\ref{gs} 
we derive the corresponding expression for the suppression factor within
the Green's function formalism and analyze its 
behavior as a function of the photon energy and the nuclear impact parameter.
In Sec.~\ref{res}, 
our model calculations within the color dipole approach based on a rigorous 
path-integral technique are first verified by comparing them with available data on 
charmonium photoproduction off proton and lead targets.
We then present predictions for 
the relative contribution of non-linear effects to the overall charmonium photoproduction cross 
section as a function of the photon energy.
The corresponding evolution rate of these non-linear
effects is subsequently compared with results based on the LO and NLO solutions of the 
BK evolution equation from Refs.~
\cite{Cepila:2025ujv,Cepila:2024qge}.
%
Finally, our findings are summarized and discussed in Sec.~\ref{conclusions}.

%
%
%
\section{Coherent quarkonium photoproduction in the Green's function formalism.
Higher-twist quark shadowing}
\label{gf}
%
%
%

In this paper, we study the non-linear QCD dynamics in the coherent photoproduction 
of quarkonia off nuclear targets utilizing the LF color dipole approach. 
This formalism has already been successfully applied to charmonium photoproduction 
off nucleons 
(see e.g., Refs.~\cite{Hufner:2000jb,Krelina:2018hmt,Cepila:2019skb})
and nuclei 
(see e.g., Refs.~\cite{Ivanov:2002kc,Nemchik:2002ug})
%
The coherent nuclear production cross section can be expressed as
\cite{Kopeliovich:2001xj},
%
\BA
  \label{sigcoh}
\sigma_{\text{coh}}^{A}(W) 
= 
\frac{1}{4}\, 
\int d^2 b \, \bigl| \mathcal{M}_{\text{coh}}(W, \vec{b}) \bigr|^2 \,,
\EA
%
where $\vec{b}$ denotes the two-dimensional nuclear impact parameter vector, and 
the production amplitude $\mathcal{M}_{\text{coh}}(W, \vec{b})$ reads
%
 \BA
\mathcal{M}_{\text{coh}}(W, \vec{b}) 
= 
\int_{-\infty}^{\infty} dz \, 
\rho_A(\vec{b}, z) \, 
\mathcal{H}(W, \vec{b}, z) \,.
%
\label{gf-coh}
 \EA
%
Here, the c.m. energy $W$ is related 
to the photon energy $k$ in the target rest frame by $W^2 = 2m_N k + m_N^2 
\approx s$, where $s = M_V \sqrt{s_N} \exp(Y)$ 
is the lepton-nucleon c.m.\ energy squared, and $Y$ 
is the rapidity variable.
The function
$\rho_A(\vec b,z)$ in Eq.~\eqref{gf-coh} represents the nuclear density distribution, 
for which we employ the realistic Wood-Saxon form with parameters taken from
Ref.~\cite{DeJager:1987qc}.

The function $\mathcal{H}(W, \vec{b}, z)$ in Eq.~\eqref{gf-coh} describes the 
mechanism of quarkonium photoproduction and can be rigorously formulated within the 
Green's function framework as,
%
 \BA
\mathcal{H}(W,\vec b,z) 
=
\int_0^1 d\alpha
\int d^{2} r_{1}\,d^{2} r_{2}\,
\Psi^{*}_{V}(\vec r_{2},\alpha)\,
G_{Q\bar Q}(z^{\prime}\to\infty,\vec r_{2};z,\vec r_{1}|\,\vec b)\,
\sigma_{Q\bar Q}(r_{1},W)\,
\Psi_{\gamma}(\vec r_{1},
\alpha)\,,
\label{f1}
 \EA
%
where $\Psi_V(\vec r,\alpha)$ and $\Psi_{\gamma}(\vec r,\alpha)$
are the LF wave function of the final-state vector meson (heavy quarkonium) 
and the LF distribution of the lowest $Q\bar Q$ Fock component of the real 
(transversely polarized) photon, respectively.
Both wave functions depend on the transverse separation $\vec{r}$ of the $Q\bar Q$ 
fluctuation (dipole) and on the boost-invariant LF fraction $\alpha = p_Q^+/p_{\gamma}^+$ carried by the heavy quark (or antiquark).
The universal dipole cross section $\sqq(r,W)$ 
describes the interaction of the $Q\bar Q$ dipole with a nucleon target,
where the energy dependence can be alternatively mapped  
via the Bjorken variable $x = M_V^2/W^2$.

The function $\mathcal{H}(W, \vec{b}, z)$ captures the spatial evolution where the 
incident real photon fluctuates coherently at the longitudinal coordinate $z$ into a colorless $Q\bar{Q}$ pair. 
This pair propagates through the nuclear medium, evolving dynamically 
until it projects onto the heavy quarkonium wave function at $z' \to \infty$. 
The Green's function $G_{Q\bar Q}(z^\prime,\vec r_2;z,\vec r_1|\,\vec b)$
describes the evolution of the interacting $Q\bar{Q}$ pair 
between the initial and final longitudinal positions $z$ and $z^\prime$ 
with corresponding transverse separations 
$\vec{r}_1$ and $\vec{r}_2$.
This function satisfies the two-dimensional Schr\"odinger equation,
\cite{Kopeliovich:1999am,Kopeliovich:2001xj}
%
 \BE
i\frac{d}{dz_2}\,G_{Q\bar Q}(z_2,\vec r_2;z_1,\vec r_1 |\,\vec b)
=
\left[\frac{m_Q^{2} - \Delta_{r_{2}}}{2\,k\,\alpha\,(1-\alpha)}
+V_{Q\bar Q}(z_2,\vec r_2,\alpha,\vec b)\right]
G_{Q\bar Q}(z_2,\vec r_2;z_1,\vec r_1 |\,\vec b)\,,
\label{schroedinger}
 \EE
%
with the boundary condition 
$G_{Q\bar Q}(z_2,\vec r_2;z_1,\vec r_1 |\,\vec b)\bigr|_{z_2=z_1} =
\delta^{(2)}(\vec r_1-\vec r_2)$. 
Here, the Laplacian $\Delta_{r_2}$ acts 
on the coordinate $r_2$ and $m_Q$ is the heavy quark mass.

The kinetic term on the right-hand side of Eq.~\eqref{schroedinger} 
regulates the relative phase shift of the propagating dipole via the longitudinal 
momentum transfer,
%
\BA
q_L(z)
=
\frac{M_{Q\bar Q}^2(z)}{2\,k}
=
\frac{m_Q^2 + q_T^2}{2\,k\,\alpha(1-\alpha)}\,,
\EA
%
where $q_T$ is the quark transverse momentum,
and the effective mass squared of the $Q\bar Q$ pair reads
%
\BA
M_{Q\bar Q}^2 = \frac{m_Q^2 + q_T^2}{\alpha (1-\alpha)}\,.
\EA
%
Consequently, taking into account a replacement $q_T^2\Longrightarrow -\Delta_r$,
the coherence length, defined as 
$l_c^{Q\bar Q} = 1/q_L$, is naturally embedded within the quantum dynamics of
Eq.~\eqref{schroedinger}.
Note, that for the heavy quarkonium photoproduction,  one can obtain 
with a rather high accuracy the coherence length given by Eq.~\eqref{lc-qq}
using a non-relativistic limit, $\alpha\sim 0.5$, $M_V^2\sim 4\, m_Q^2$ and $m_Q^2\gg q_T^2$.

The real part of the complex light-front potential 
$V_{Q\bar{Q}}(z_2, \vec{r}_2, \alpha, \vec{b})$ 
in Eq.~\eqref{schroedinger} governs the non-perturbative 
binding forces between the $Q$ and $\bar{Q}$, whereas 
its imaginary part accounts for the absorptive attenuation 
of the $Q\bar{Q}$ pair in the nuclear medium. 
For the propagation of $Q\bar Q$ fluctuations in a vacuum, 
the potential $V_{Q\bar Q}$ reduces to its real part only. 
To determine the properties of these vacuum $Q\bar Q$ fluctuations, 
we follow the established framework of 
Refs.~\cite{Ivanov:2002kc,Nemchik:2024lny,Nemchik:2025myg}.
%
This method is based on well-defined quarkonium wave functions, 
which are generally obtained by solving the non-relativistic 
Schr\"odinger equation in the $Q\bar Q$ rest frame with realistic potentials. 
In this work, we adopt the specific power-like (POW) potential
\cite{Martin:1980jx,Barik:1980ai},
since it provides the best description of available data 
on proton targets, as demonstrated in 
Ref.~\cite{Cepila:2019skb}.

To transform these rest-frame wave functions into the LF frame, we employ the standard Terent'ev prescription
\cite{Terentev:1976jk}
detailed in 
Refs.~\cite{Hufner:2000jb,Krelina:2018hmt,Cepila:2019skb,Krelina:2019egg}
(see also Sec.~II.B in Ref.~\cite{Kopeliovich:2022jwe}). 
The theoretical justification of this boosting procedure 
for heavy $Q\bar Q$ bound states was established in 
Ref.~\cite{Kopeliovich:2015qna}
by utilizing a Lorentz-boosted Schr\"odinger equation 
derived from the Lorentz-invariant Bethe-Salpeter equation.
During the Terent'ev transformation, the transverse motion 
of the quarks prevents their transverse momenta from 
aligning parallel to the boost axis. 
Following 
Refs.~\cite{Krelina:2018hmt,Lappi:2020ufv,Kopeliovich:2022jwe},
we consistently include the essential kinematic correction
\cite{Hufner:2000jb}
arising from the Melosh spin rotation effects
\cite{Melosh:1974cu}.

Finally, following the formal procedure introduced in 
Ref.~\cite{Nemchik:2024lny},
the real part of the LF potential, 
${\mathcal Re} V_{Q\bar Q}$, is uniquely reconstructed from 
the rest-frame interaction potential. 
This step ensures that the dynamically generated 
LF potential remains strictly consistent with the 
structure of the boosted quarkonium wave functions 
described above.

Considering the propagation of an interacting $Q\bar Q$ pair 
in the nuclear medium, the LF potential 
$V_{Q\bar Q}$ in Eq.~\eqref{schroedinger} incorporates 
an imaginary part in addition to the real part and 
takes the following form,
%
 \BE
{\mathcal Im} V_{Q\bar Q}(z_2,\vec r,\alpha,\vec b) 
= 
- \frac{\sigma_{Q\bar Q}(\vec r,W)}{2}\,\rho_{A}({b},z_2)\,.
\label{pot-im}
 \EE 
%
The validity of this absorption potential has been verified through calculations of nuclear shadowing in deep-inelastic scattering off nuclei, demonstrating good agreement with experimental data, as shown in 
Refs.~\cite{Kopeliovich:1998gv,Kopeliovich:2000ra,Nemchik:2003wx,Kopeliovich:2008ek,Krelina:2020ipn}.
%
In this work, we implement the modern Golec--Biernat--Sapeta (GBS) saturated parametrization
\cite{Golec-Biernat:2017lfv}
for the dipole cross section: 
$\sigma_{Q\bar Q}(r,W) = \sigma_0 \,\bigl( 1 - \exp[ - Q_s^2(W)\,r^2/4] \bigr)$, where
$Q_s^2 = Q_0^2 (W/W_0)^{2\lambda}$, $W_0 = M_V/\sqrt{x_0}$, and the model parameters are given by
$\sigma_0 = 27.32\text{ mb}$, $x_0 = 0.42\cdot 10^{-4}$, and $\lambda = 0.248$.

To improve the precision of our calculations,
we incorporate a small correction in Eq.~\eqref{sigcoh} 
due to the real part of the $\gamma^*~N\to V~N$ amplitude  
applying the following replacement 
\cite{Bronzan:1974jh,Nemchik:1996cw,Forshaw:2003ki},
%
\BA
\sqq(r,W)
\Rightarrow
\sqq(r,W)\cdot\left(1 - i\,\frac{\pi}{2}\,\Lambda \right),
\qquad\qquad
\Lambda = 
\frac{1}{2}\,\frac{\partial\,\ln\,{\sqq(r,W)}}{\partial\,\ln W}\, .
  \label{re/im}
\EA
%
%
%
Furthermore, the skewness correction originating from 
the non-forward kinematics of gluon exchange 
\cite{Shuvaev:1999ce} 
is accounted for via the factor $R_S$:
%
\BA
\sqq(r,W)
&\Rightarrow&
\sqq(r,W)\cdot R_S(\Lambda)\,,
\qquad\qquad 
R_S(\Lambda)
=
\bigl ({2^{2\,\Lambda + 3}}/{\sqrt{\pi}}\, \bigr) \,\cdot 
{\Gamma(\Lambda + 5/2)}/{\Gamma(\Lambda + 4)}\,.
  \label{skewness-a}
\EA
%

Due to abnormally large weight of a $D$-wave component in quarkonium wave functions, as analyzed in 
Ref.~\cite{Kopeliovich:2022jwe},
we eliminated the standard photon-like vertex for a heavy 
quarkonium transition $V\to Q\bar Q$ and thus adopted 
the $V\to Q\bar Q$ structure from our previous studies
\cite{Ivanov:2002kc,Ivanov:2007ms,Krelina:2018hmt,Cepila:2019skb,Kopeliovich:2022jwe}.
%
Then, assuming $s$-channel helicity conservation, the function 
$\mathcal{H}(W,\vec b,z)$ in Eq.~\eqref{gf-coh} has the following form,
%
\BA
\mathcal{H}(W,\vec b,z) 
=
\int_0^1 d\alpha\int d^{2} r_{1}\,d^{2} r_{2}\,
G_{Q\bar Q}(z^\prime,\vec r_{2};z,\vec r_{1}|\,\vec b)
\Bigl|_{z^\prime\to\infty}\,\cdot\,
\Omega(r_1,r_2,\alpha)\cdot\sqq(r_{1},W)\, ,
\label{f1-sr}
\EA
%
where $\Omega(r_1,r_2,\alpha) 
= \Sigma^{(1)}(r_1,r_2,\alpha) + \Sigma^{(2)}(r_1,r_2,\alpha)$.
In charmonium photoproduction, one can safely adopt with a rather 
good accuracy the perturbative wave functions for the incoming photon, 
$\Psi_{\gamma}(\vec r,\alpha) = K_0(\eta\,r)$ with $\eta = m_Q$. 
Then the functions $\Sigma^{(1,2)}$ read
%
\BA
  \Sigma^{(1,2)}(r_1,r_2,\alpha)
   =  
   N\,
   K_{0,1}(\eta\,r_1) \int\limits_{0}^{\infty} dp_T\,p_T\,
   J_{0,1}(p_T r_2) \Psi_V (p_T,\alpha) \,
    \mathcal{R}^{(1,2)}(p_T)
\label{sigma-sr1}
\EA
%
with
%
\BA
\mathcal{R}^{(1)}(p_T) 
=
\frac{2\,m_Q^2(m_L+m_T)+m_L\,p_T^2} { m_T (m_L + m_T)} \,,
\qquad\qquad\quad
\mathcal{R}^{(2)}(p_T) 
=
\frac{m_Q^2(m_L+2m_T)-m_T\,m_L^2} {m_Q^2\,m_T (m_L+m_T)}\, 
\eta\,p_T\,.
\label{sigma-sr2}
\EA
%
Here $N=Z_Q\,\sqrt{2 N_c \,\aem} /4\,\pi$, 
where the factor $N_c=3$ represents the number of colors in QCD, 
$Z_Q=\frac{2}{3}$ is the charge-isospin factor for the production of charmonia,  
and $J_{0,1}$ are the Bessel functions of the first kind.
The variables $m_{T,L}$ in Eq.~\eqref{sigma-sr2} 
have the following form,
%
\BE
m_T = \sqrt{m_Q^2 + p_T^2} \,, 
\qquad m_L = 2\, m_Q\,\sqrt{\alpha(1-\alpha)}\,.
\EE
%
The value of the heavy quark mass $m_Q$ is consistently fixed by the choice of the POW potential parametrization used throughout our numerical calculations.

In the high-energy asymptotic limit where the coherence length 
vastly exceeds the nuclear size 
($l_c^{Q\bar{Q}} \gg R_A$, corresponding to the long coherence length 
(LCL) regime), 
the Green's function acquires a simple form
%
 \BE
G_{Q\bar Q}(z_2,\vec r_2;z_1,\vec r_1 |\,\vec{b}) 
\Longrightarrow
\delta^{(2)}(\vec r_1-\vec r_2)\,\exp\left[
- \frac{1}{2}\,\sigma_{Q\bar Q}(r_1,s)
\int\limits_{z_1}^{z_2} dz\,\rho_A(\vec b,z)\right]\, .
\label{gf-lcl}
 \EE
%
This is the well-known ``frozen'' approximation, where 
the transverse size of the $Q\bar{Q}$ fluctuation 
is preserved during its propagation through the nucleus 
due to Lorentz time dilation.
Under this condition, the production amplitude 
in Eq.~\eqref{gf-coh} evaluates to
%
 \BE
{{\cal M}}_{\text{coh}}(W,\vec b)\Bigr|_{l_c^{Q\bar Q} \gg R_A} 
=
2\,\int d^2r\int_0^1 d\alpha\,
\Psi_{V}^{*}(\vec r,\alpha)\,
\Biggl (1 -\exp\left[-\frac{1}{2}\,\sqq(r,W)\,T_A(\vec b)\right]\Biggr )\,
\Psi_{\gamma}(\vec r,\alpha)\, ,
\label{coh-lcl}
 \EE
 %
where $T_A(\vec b) = \int_{-\infty}^{\infty} dz \,\rho_A(\vec b,z)$ represents the nuclear thickness function,
normalized such as $\int_{0}^{\infty} d^2 b~T_A(\vec b) = A$, where $A$ is the nuclear mass number.

%
%
%
\section{LEADING-TWIST GLUON SHADOWING}
\label{gs}
%
%
%

In the previous section, we presented the higher-twist quark shadowing in the 
photoproduction of heavy quarkonia off nuclei, which corresponds to the lowest 
$|Q\bar{Q}\rangle$ Fock component of the photon. 
Its magnitude depends on the 
onset of quantum coherence effects, which are naturally incorporated within the 
Green's function formalism. 
Here, the primary condition for the maximal strength 
of this phenomenon is a sufficiently long coherence length, $l_c^{Q\bar{Q}} \gtrsim R_A$. 
However, at higher photon energies where 
$l_c^G \gsim d\approx 2\,\fm$, 
one must also include shadowing corrections originating 
from higher Fock components of the photon containing gluons, i.e., 
$|Q\bar{Q}G\rangle, |Q\bar{Q}2G\rangle, \dots, |Q\bar{Q}nG\rangle$.
Such corrections are widely known as leading-twist 
gluon shadowing, as introduced in 
Ref.~\cite{Kopeliovich:1999am} 
within the QCD dipole representation.
In our calculations, the GS correction is incorporated
via the following replacement,
\cite{Kopeliovich:2001ee},
%
\BE
  \sqq(r,W) \Rightarrow \sqq(r,W) \cdot R_G(W,\vec b)\,,
\label{eq:dipole:gs:replace}
\EE
%
which leads to a reduction of the dipole cross section $\sqq(r, W)$ in 
nuclear reactions relative to processes on a free nucleon. 
Although rigorous calculations within the Green's function formalism
\cite{Kopeliovich:1999am,Kopeliovich:2001xj,Kopeliovich:2001hf,Ivanov:2002kc}
allow an explicit evaluation of the correction factor 
$R_G(W, \vec{b})$ only for the lowest gluonic 
$|Q\bar{Q}G\rangle$ component of the photon, the effects 
of higher multi-gluon Fock states can be effectively 
taken into account by eikonalizing the calculated 
$R_G(W, \vec{b})$, as argued in 
Ref.~\cite{Kopeliovich:2001ee}.

The gluon shadowing phenomenon admits different interpretations 
depending on the chosen reference frame. 
In the infinite-momentum frame of the nucleus, 
it can be treated as glue-glue fusion.
While the Lorentz contraction of the nucleus and 
the bound nucleons preserves their structural separation, 
the gluon clouds of nucleons at small momentum 
fractions $x$ are less contracted.
This causes them to overlap and interact at small $x$, 
generating a fusion process of gluons originating from different 
nucleons, which subsequently reduces the nuclear gluon density.

Conversely, in the rest frame of the nucleus, the same process 
is interpreted as a consequence of the eikonalization 
of the higher Fock components of the incident particle. 
The standard eikonalization given by Eq.~\eqref{coh-lcl} 
corresponds to the lowest $|Q\bar{Q}\rangle$ Fock state 
of the photon without the GS correction. 
Equation~\eqref{coh-lcl} implicitly assumes that none 
of the radiated gluons participate in multiple interactions 
within the nucleus, even though the phenomenological 
dipole cross section $\sigma_{Q\bar{Q}}(r, W)$ implicitly 
includes all vacuum effects of gluon radiation. 
The leading-order correction related to GS is then generated by the 
eikonalization of the next Fock component, namely the 
$|Q\bar{Q}G\rangle$ state.

In contrast to the long coherence length approximation 
utilized for the higher-twist quark shadowing 
at high photon energies ($l_c^{Q\bar{Q}} \gg R_A$), 
we take into account 
the $|Q\bar{Q}G\rangle$ component of the photon 
without any kinematic restrictions on $l_c^G$ 
by employing the light-front Green's function formalism
\cite{Kopeliovich:1999am}.
%
For our evaluation of the factor $R_G$, we have introduced several improvements over 
the calculations in 
Ref.~\cite{Ivanov:2002kc}.
%
Specifically, our approach incorporates the following refinements:
\\
i) \textit{Massive quarks:} We introduce non-zero quark masses instead of relying on the massless approximation. \\
ii) \textit{Effective gluon mass variation:} We systematically analyze 
the impact of different values for the effective gluon mass, 
which acts as an infrared regulator. \\
iii) \textit{GBS dipole cross section:} The calculation is carried out using 
the modern GBS dipole cross section model, which extends the original GBW 
framework by incorporating Dokshitzer--Gribov--Lipatov--Altarelli--Parisi
evolution and utilizing the final combined HERA data. \\
%
%
The corresponding 
factor $R_G$ is related to the shadowing for longitudinal virtual photons via
%
\BE
    R_G(x,Q^2) 
    = 
    \frac{G_A(x,Q^2)}{A\,G_N(x,Q^2)} 
    \approx 
    \frac{\sigma_A^L(x,Q^2)}{A\,\sigma_N^L(x,Q^2)}
    \approx 
    1 - \frac{\Delta \sigma_{Q\bar{Q}G} (x, Q^2)}
    {A\,\sigma_{L,tot}^{\gamma^*N}(x,Q^2)}\,,
\label{RG-1}
\EE
%
where the variable $x$ is related to the c.m. photon-nucleon energy $W$ by
%
\BE
x = \frac{M_V^2 + Q^2}{W^2 + Q^2 - m_N^2}\,.
\EE
%
When the integration over the nuclear impact parameter is not performed, one obtains 
the $\vec{b}$-dependent gluon shadowing factor:
%
\BE
\label{RG-2}
    R_G (\vec b, x, Q^2) 
    =
    1-\frac{\Delta \sigma_{Q\bar{Q}G} (\vec b, x, Q^2)}
    {T_A(\vec b)\,\sigma_{L,tot}^{\gamma^*N}(x, Q^2)}\,,
\EE
%
where $\int d^2b \, \Delta\sigma_{Q\bar{Q}G}(\vec{b}, x, Q^2) = \Delta\sigma_{Q\bar{Q}G}(x, Q^2)$.

The total longitudinal photoabsorption cross section $\sigma_{\text{L,tot}}^{\gamma^* N}(x, Q^2)$ in the denominator of Eqs.~\eqref{RG-1} and~\eqref{RG-2} reads,
%
\BA
\label{stot-L}
    \sigma_{L,tot}^{\gamma^* N}(x,Q^2) 
    &=&
    \sum_{f} \int d^2 r\int_0^1 d\alpha\,
    \Bigl | \Psi_{\gamma^*\rightarrow f\bar{f}}^L (r,\alpha,\widetilde{Q}^2) 
    \Bigr |^2\,\sigma_{Q\bar{Q}}(r,x) 
    \nonumber 
    \\
    &=&
    8\,\widetilde{Q}^2\frac{\aem\, N_C}{4\pi^2}
    \Big(\sum_{f} Z_f^2 \int_0^1 d\alpha\,\alpha^2(1-\alpha)^2 
    \int d^2r\, K_0^2(\epsilon_f \,r)\,\sigma_{Q\bar{Q}}(r, x)\Big)\,, 
\EA
%
where $\aem \approx 1/137$ is the fine structure constant, 
$\widetilde{Q}^2 = Q^2+M_V^2$,
$N_C = 3$ is the number of colors in QCD, $\epsilon_f^2 = m_f^2+\alpha(1-\alpha)\,\widetilde{Q}^2$, 
$m_f$ and $Z_f$ are the mass and charge of the quark of flavor $f$, respectively.
For small $Q\bar{Q}$ transverse separations $\sim 1/\sqrt{M_V^2 + Q^2}$, 
one can implement the quadratic approximation 
of the dipole cross section, $\sqq(r, x) = C_{Q\bar{Q}}(x) r^2$, where 
$C_{Q\bar{Q}}(x) = [d\sigma_{Q\bar{Q}}(r, x) / dr^2]_{r \to 0}$. 
This allows a 
simplification of Eq.~\eqref{stot-L} by integrating explicitly over $r$:
%
\BA
    \int d^2r\, K_0^2(\epsilon_f \,r)\, 
    \sigma_{Q\bar{Q}}(r, x) 
    = 
    \int2\pi \,r\, d r\, K_0^2(\epsilon_f \,r)\, C_{Q\bar{Q}}(x)\,r^2 
    =    
    2\pi\,C_{Q\bar{Q}}(x) \int\,d r\,r^3 K_0^2(\epsilon_f \,r) 
    = 
    \frac{2\pi\,C_{Q\bar{Q}}(x)}{3\,\epsilon_f^4}\,.
\EA
%
The total longitudinal $\gamma^* N$ cross section then reduces to
%
\BA
    \sigma_{L,tot}^{\gamma^*N}(x,Q^2)
    &=&
    8\, \widetilde{Q}^2\,C_{Q\bar{Q}}(x)\,\frac{\aem\,N_C}{2\pi} 
    \Big(\sum_{f} Z_f^2\int_0^1 d \alpha \, \frac{\alpha^2(1-\alpha)^2}{3\,\epsilon_f^4}\Big) 
    \nonumber\\
    &=& 
    \frac{4\,C_{Q\bar{Q}}(x)\,\aem\,N_C}{3\pi} \Big(\sum_{f} Z_f^2\int_0^1 d\alpha \, \frac{\alpha^2(1-\alpha)^2\,\widetilde{Q}^2}{(m_f^2+\alpha(1-\alpha)\, \widetilde{Q}^2)^2}\Big)\,. 
    \label{eq:nucleon_c-s_final}
\EA
%

The inelastic correction  $\Delta \sigma_{Q\bar{Q}G} (x, Q^2)$ to the total nuclear photoabsorption cross section $\sigma_A^L(x,Q^2)$ reads
\cite{Kopeliovich:1999am},
%
\BA
  \Delta\sigma_{Q\bar QG}(x,Q^2) 
  &=&
  8\pi\,{\mathcal Re}\int\!d^2b \int\!dM^2\,
      \left.\frac{d^2\sigma(\gamma^*N\to Q\bar QG N)}{dM^2\,dq_T^2}    
      \right|_{q_T=0}
  \label{qqg}\\
  &\times&
  \int\limits_{-\infty}^{\infty}\!\!dz_1
  \int\limits_{-\infty}^{\infty}\!\!dz_2\,
    \Theta(z_2-z_1)\,\rho_A(\vec{b},z_1)\,\rho_A(\vec{b},z_2)\,
    \exp\left[-i\,q_L\,(z_2-z_1)\right]\,,
  \nonumber 
\EA
%
where the invariant mass squared of the three-body fluctuation is given by
%
\BE
  M^2\equiv M^2_{Q\bar QG} 
  =\sum\limits_i\frac{m_i^2+k_i^2}{\alpha_i}\,,
  \label{M2qqg}
\EE
%
with the sum running over partons ($i=Q, \bar{Q}, G$) carrying masses $m_i$, 
transverse momenta $\vec{k}_i$, and light-front momentum fractions $\alpha_i$. 
The diffractive production of the $Q\bar{Q}G$ system described by Eq.~\eqref{qqg} 
corresponds to an intermediate state in a double scattering process within the nucleus.
Here, $z_1$ and $z_2$ denote the longitudinal coordinates of the target nucleons $N_1$ 
and $N_2$ participating in the diffractive transition $\gamma^*\,N_1 \to Q\bar{Q}G\,N_1$ 
and its subsequent inverse transition $Q\bar{Q}G\,N_2 \to \gamma^*\,N_2$. 
The magnitude of 
this shadowing correction is controlled by the longitudinal momentum transfer
%
\BE
  q_L=\frac{Q^2+M^2}{2\,k}~,
  \label{ql-lcg}
\EE
%
which is related to the coherence length of the $|Q\bar{Q}G\ra$
Fock state of the photon by $l_c^G = 1/q_L$.

Within the path-integral technique, Eq.~\eqref{qqg} can be expressed as,
%
\BA
\label{dsigma-qqg}
        \Delta \sigma_{Q\bar{Q}G} (\vec b,x,Q^2)
        = 
        \frac{1}{2}\,{\mathcal Re}\int_{-\infty}^{+\infty} d z_2
        \int_{-\infty}^{z_2} d z_1\,\rho_A(\vec b, z_1)\,\rho_A(\vec b, z_2)
        \int d^2r_1\,d^2\rho_1\,d^2r_2\,d^2\rho_2 
        \nonumber
        \\
        \times\int d\alpha 
        \int\limits_x^{x_{max}} \frac{d\alpha_G}{\alpha_G}\,
        F^\dagger_{\gamma\rightarrow Q\bar{Q}G}(\vec{r}_2,\vec{\rho}_2,\alpha, \alpha_G)\,
        G_{Q\bar{Q}G} (\vec{r}_2, \vec{\rho}_2, z_2; \vec{r}_1,\vec{\rho}_1, z_1\,|\,\vec b)\,
        F_{\gamma\rightarrow Q\bar{Q}G} (\vec{r}_1, \vec{\rho}_1, \alpha, \alpha_G) 
        \,,
\EA
%
where $\vec{r}_{1,2}$ are the transverse separations between the $Q$ and $\bar{Q}$ at 
longitudinal positions $z_1$ and $z_2$, respectively, and $\vec{\rho}_{1,2}$ denote the 
transverse distances from the gluon to the center of mass of the $Q\bar{Q}$ pair. 
The light-front fraction carried by the gluon $\alpha_G$ is bounded from above by $x_{\text{max}} = 0.1$. 
The function $F_{\gamma \to Q\bar{Q}G}$ is the light-front function for the 
$\gamma^* \to Q\bar{Q}G$ transition, and $G_{Q\bar{Q}G}$ represents the three-body Green's 
function describing the simultaneous propagation and mutual interaction of the 
$Q\bar{Q}G$ system inside the nuclear medium.

The three-body Green's function can be simplified, assuming $\alpha_G\ll 1$ and 
sufficiently smaller $Q$-$\bar Q$ separation compared to 
the gluon propagation radius $r_G\sim 0.25\text{--} 0.30\,\fm$ 
\cite{Kopeliovich:1999am,Kopeliovich:2007pq}.
In this case we arrive at a following factorized structure,
%
\BA
\label{factorisation}
    G_{Q\bar{Q}G} (\vec{r}_2, \vec{\rho}_2, z_2; \vec{r}_1,\vec{\rho}_1, z_1\,|\,\vec b) 
    = 
    G_{Q\bar{Q}}(\vec{r}_2, z_2; \vec{r}_1, z_1\,|\,\vec b)\,\, 
    G_{GG}(\vec{\rho}_2, z_2; \vec{\rho}_1, z_1\,|\,\vec b)\,,
\EA
%
where 
$G_{GG}(\vec\rho_2,z_2;\vec\rho_1,z_1\,|\,\vec b)$ describes the propagation of the
$GG$ dipole (in fact the color-octet $Q\bar Q$ and gluon) in the nuclear
medium. 
This Green's function $G_{GG}$ satisfies the following two dimensional Schr\"odinger
equation analogous to Eq.~\eqref{schroedinger}, but
incorporating now the glue-glue LF potential $V_{GG}(\vec\rho,z)$,
%
\BE
    i\frac{d}{dz_2} G_{GG} (\vec{\rho}_2, z_2; \vec{\rho}_1, z_1\,|\,\vec b) 
    = 
    \Bigg[-\frac{\Delta_{\vec{\rho}_2}}{2\,k\,\alpha_G (1-\alpha_G)} 
    +V_{GG} (\vec{\rho}_2, z_2)\Bigg]\,
    G_{GG} (\vec{\rho}_2, z_2; \vec{\rho}_1, z_1\,|\,\vec b)\,,
    \label{schrodinger-gg}
\EE
%
with the boundary condition $G_{GG} (\vec{\rho}_2, z_2; \vec{\rho}_1, z_1\,|\,\vec b)\big|_{z_2 \,=\,z_1} = \delta^{(2)}(\vec{\rho}_2 \,-\, \vec{\rho}_1)$. 
The absorptive imaginary part of this complex glue-glue potential scales with the nuclear density:
%
\begin{equation}
    {\mathcal Im} \,V_{GG} (\vec{\rho}, z) 
    = 
    - \frac{1}{2}\sigma_{GG}(\vec{\rho})\,\rho_A(b, z)\,,
\label{Im-Vgg}
\end{equation}
%
where the glue-glue dipole cross section is related to the $\sqq$
by the relation,
%
\BE
  \sigma_{GG}(r,x)
  =
  \frac{9}{4}\,\sqq(r,x)\,.
  \label{sgg-sqq}
\EE
%
Following 
Ref.~\cite{Kopeliovich:1999am},
%
we assume a quadratic shape for the real part of the glue-glue potential,
%
\BE
  {\mathcal Re}\,V_{GG}(\vec \rho,z) =
  \frac{b_0^4\,\rho^2}{2\,k\,\alpha_G(1-\alpha_G)}\,,
  \label{Re-Vgg}
\EE
%
where the parameter $b_0 = 0.65,\GeV$ is inversely related to the gluon propagation radius
($b_0 \sim 1/r_G$) and is strongly supported by 
the data on diffractive gluon
radiation (the triple-Pomeron contribution in terms of Regge approach)
which is an essential part of Gribov's inelastic shadowing 
\cite{Gribov:1968jf}.
%
The interpretation of such rather large value of $b_0$ is based on the well known 
smallness of diffractive cross section.
This leads to a rather weak gluon shadowing since strong interaction squeezes 
the glue-glue wave packet resulting in small nuclear attenuation due to color transparency.

As the next step we implement the quadratic approximation for the dipole cross-section, 
valid at small glue-glue transverse separations, $\sigma_{GG}(r,x) = C_{GG}(x)\, r^2$,
with $C_{GG}(x) = d\sigma_{GG}(r,x)/dr^2\Bigl |_{r=0}$.
In this case, both the real and imaginary parts of the glue-glue LF potential resemble a harmonic oscillator potential.
This allows an exact analytical solution for
the Green's function $G_{GG}$ in Eq.~\eqref{schrodinger-gg}.
Performing the spacial integration 
in Eq.~\eqref{dsigma-qqg}, we arrive at the following
expression for the shadowing correction
\cite{Kopeliovich:1999am,Ivanov:2002kc},
%
\begin{equation}
\label{dsigma-qqg-1}
    \begin{split}
        \Delta \sigma_{Q\bar{Q}G} (b, x, Q^2)= \int\limits_{-\infty}^{+\infty} dz_2\int\limits_{-\infty}^{z_2}dz_1 \rho_A(b, z_1) \rho_A(b, z_2) \int\limits_x^{x_{max}} \frac{d\alpha_G}{\alpha_G}
        \frac{16\,\aem\big(\sum_f Z_f^2\big)\,\alpha_{S}(\widetilde{Q}^2)\,C^2_{GG}(\tilde{x}, \widetilde{Q}^2)}{3\pi ^2\widetilde{Q}^2\tilde{b}^2} F(\tilde{x}, 
        \widetilde{Q}^2, \alpha_G, \Delta z)~,
        \end{split}
\end{equation}
%
where
%
\begin{gather}
    F(\tilde{x}, \widetilde{Q}^2, \alpha_G, \Delta z) = {\mathcal Re}\Bigg[\mathrm{exp}\Bigg(-\frac{\Omega \Delta z}{2t}\Bigg)
    \Bigg\{\frac{t}{w} + \frac{\mathrm{sinh}(\Omega \Delta z)}{t}\mathrm{ln}\Bigg(1-\frac{t^2}{u^2}\Bigg) + \frac{2t^3}{uw^2}+\frac{t\,\mathrm{sinh}(\Omega \Delta z)}{w^2} + \frac{4t^3}{w^3} \Bigg\}\Bigg]\,,\\
    \Delta z = z_2 - z_1\,,\\
     \tilde{x}= \mathrm{min}(x/\alpha_G, 0.1)\,,\\
    \alpha_S(\widetilde{Q}^2) = \frac{4\pi}{9\,\mathrm{ln}[(\widetilde{Q}^2+0.25 \mathrm{GeV}^2)/\Lambda_{QCD}^2]},\,\,\,\,\,\,\Lambda_{QCD}\approx200\,\mathrm{MeV} \,,\\
    B = \sqrt{\tilde{b}^4-i\alpha_G(1-\alpha_G)\, k\, \rho_A C_{GG}(\tilde{x}, \widetilde{Q}^2)}\,, \\
    \Omega = \frac{iB}{\alpha_G(1-\alpha_G)\,k} \,,\\
    t = \frac{B}{\tilde{b}^2} \,,\\
    u = t \,\mathrm{cosh}(\Omega \Delta z) + \mathrm{sinh}(\Omega \Delta z) \,,\\
    w =  (1+t^2) \,\mathrm{sinh}(\Omega \Delta z) + 2t \,\mathrm{cosh}(\Omega \Delta z) \,.  
\end{gather}
%

The factor $C_{GG}(x,Q^2)$ in Eq.~\eqref{dsigma-qqg-1} is determined from the condition of equivalent descriptions of shadowing in the limit of large $l_c^G \gg R_A$ using the realistic parametrization for $\sigma_{GG}(r,x)$ and its simplified form 
$\sigma_{GG}(r,x) = C_{GG}(x)\, r^2$,
%
\begin{equation}
    \begin{split}
        \frac{\int d^2b \int d^2r \big|\Psi_{QG}(r, \alpha_G)\big|^2\big[1-\mathrm{exp}\big(-\frac{9}{8} \sigma_{Q\bar{Q}}(r, x) T_A(b)\big)\big]}{\int d^2r \big|\Psi_{QG}(r, \alpha_G)\big|^2\frac{9}{4} \sigma_{Q\bar{Q}}(r, x)} = \\
        =\frac{\int d^2b \int d^2r \big|\Psi_{QG}(r, \alpha_G)\big|^2\big[1-\mathrm{exp}\big(-\frac{1}{2} C_{GG}(x,Q^2)\,r^2 \,T_A(b)\big)\big]}{\int d^2r \big|\Psi_{QG}(r, \alpha_G)\big|^2 C_{GG}(x,Q^2)\,r^2}\,,
    \end{split}
\end{equation}
%
with the non-perturbative LF wavefunction of the $GG$ dipole of the form,
%
\begin{equation}
    \big|\Psi_{QG}(r, \alpha_G)\big|^2 = \frac{4\alpha_S(\widetilde{Q}^2)}{3\pi^2} \frac{\mathrm{exp}(-\tilde{b}^2r^2)}{r^2}\,,
\end{equation}
%
where $\alpha_S$ is the running coupling constant, 
and $\tilde{b}^2 = 1 / r_G^2 + \alpha_G\widetilde{Q}^2$.

For heavy nuclei we can safely use with rather good accuracy
the approximation of constant nuclear density $\rho_A(r)=\rho_0\,\Theta(R_A-r)$.
Consequently, the nuclear thickness function $T_A(b)=\rho_0\,L(b)$,
where $L(b)=2\sqrt{R_A^2-b^2}$. 
This allows to perform the integration in Eq.~\eqref{dsigma-qqg-1} over coordinates $z_{1,2}$.
Then after substitution of the photoabsorption cross section on the nucleon,
Eq.~\eqref{eq:nucleon_c-s_final}, and 
the gluon shadowing correction, Eq.~\eqref{dsigma-qqg-1}, into Eq. \eqref{RG-2},
we obtain the final expression 
for the $b$-dependent gluon shadowing factor in the following form,
%
\begin{equation}
\label{eq:gs_factor_bdep_final}
\begin{split}
    R_G(b, x, Q^2) 
    = 
    1 - \frac{\rho_0^2}{T_A(b)}  \int\limits_0^{L(b)} dz\,
    [L(b) - z]\int\limits_x^{x_{max}} 
    \frac{d\alpha_G}{\alpha_G} 
    \,\frac{4\,\alpha_S(\widetilde{Q}^2)\,C^2_{GG}(\tilde{x}, 
    \widetilde{Q}^2)}
    {3\,\pi\,\tilde{b}^2\,C_{Q\bar{Q}}(x)}\,
    \mathcal{M}(\widetilde{Q}^2, f)\, 
    F(\tilde{x}, \widetilde{Q}^2, \alpha_G, z)\,,
\end{split}
\end{equation}
%
where the factor $\mathcal{M}(\widetilde{Q}^2, f)$ accounts for the masses of the active quark flavors $f$ and reads,
%
\begin{equation}\label{eq:quark_mass_term}
    \mathcal{M}(\widetilde{Q}^2, f) 
    = 
    \frac{\sum\limits_{f} Z^2_f }
    { \widetilde{Q}^4\,\sum\limits_{f} \Big[Z^2_{f} \int\limits_0^1 
    d\alpha\,\Big ( \frac{\alpha (1-\alpha)}    
    {m_{f}^2 + \widetilde{Q}^2\alpha(1-\alpha)}\Big )^2\, \Big]}\,.
\end{equation}
%
Here, in photoproduction of heavy quarkonia, this factor is close to unity due to strong inequalities, $\widetilde{Q}^2\gg m_f^2$. 
Nevertheless, in this paper we took the effective quark masses 
$m_{u,d} = 0.3\,\GeV$ and $m_s = 0.45\,\GeV$ following
Ref.~\cite{Kopeliovich:1998gv}.
Such a choice 
provides a more physically grounded description that slightly enhances the gluon shadowing effects compared to the massless approximation, $\mathcal{M}(\widetilde{Q}^2, f) = 1$.

\BF
%
\includegraphics[height=6.5cm]{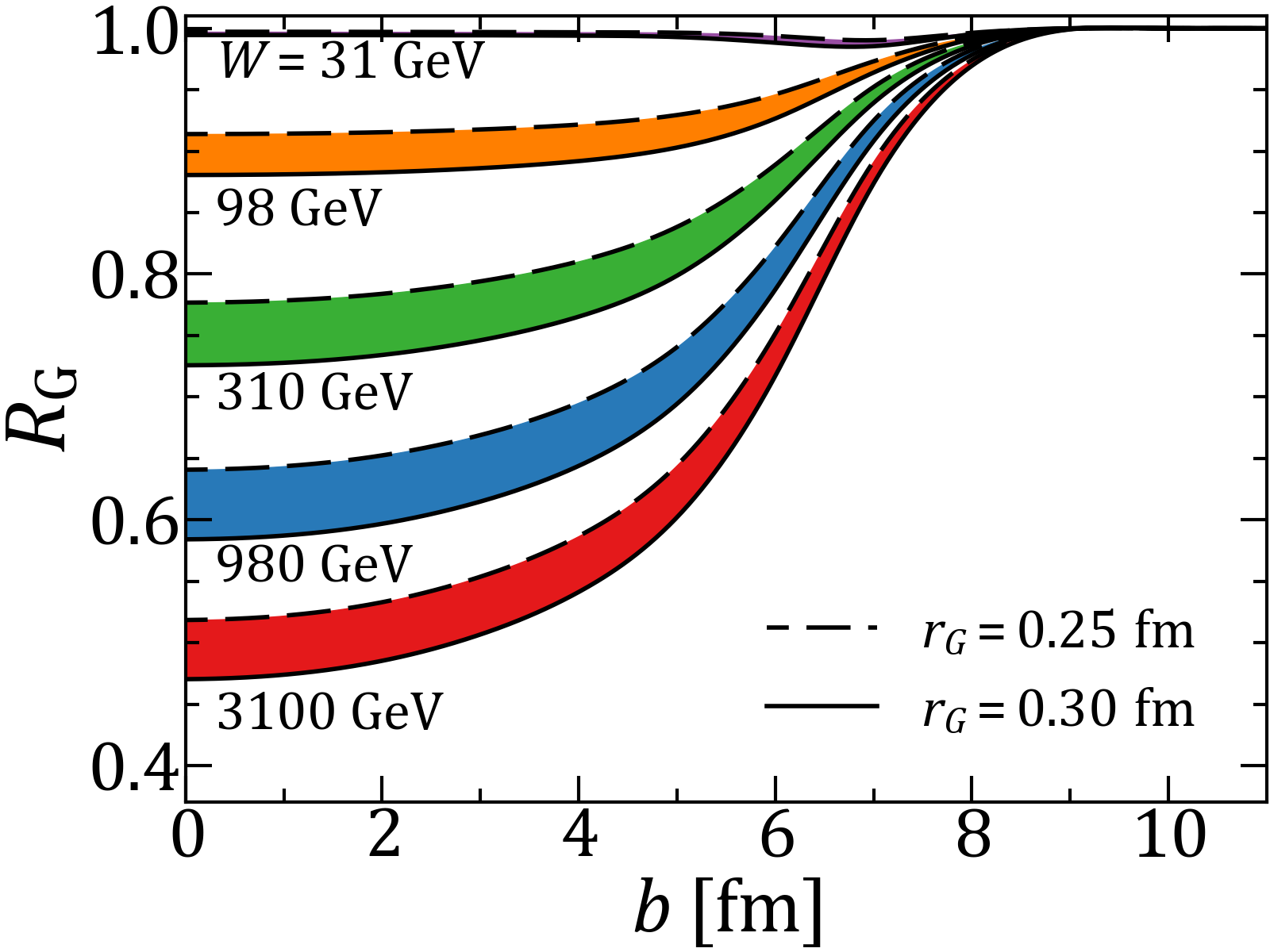}
\hspace*{0.10cm}
\includegraphics[height=6.5cm]{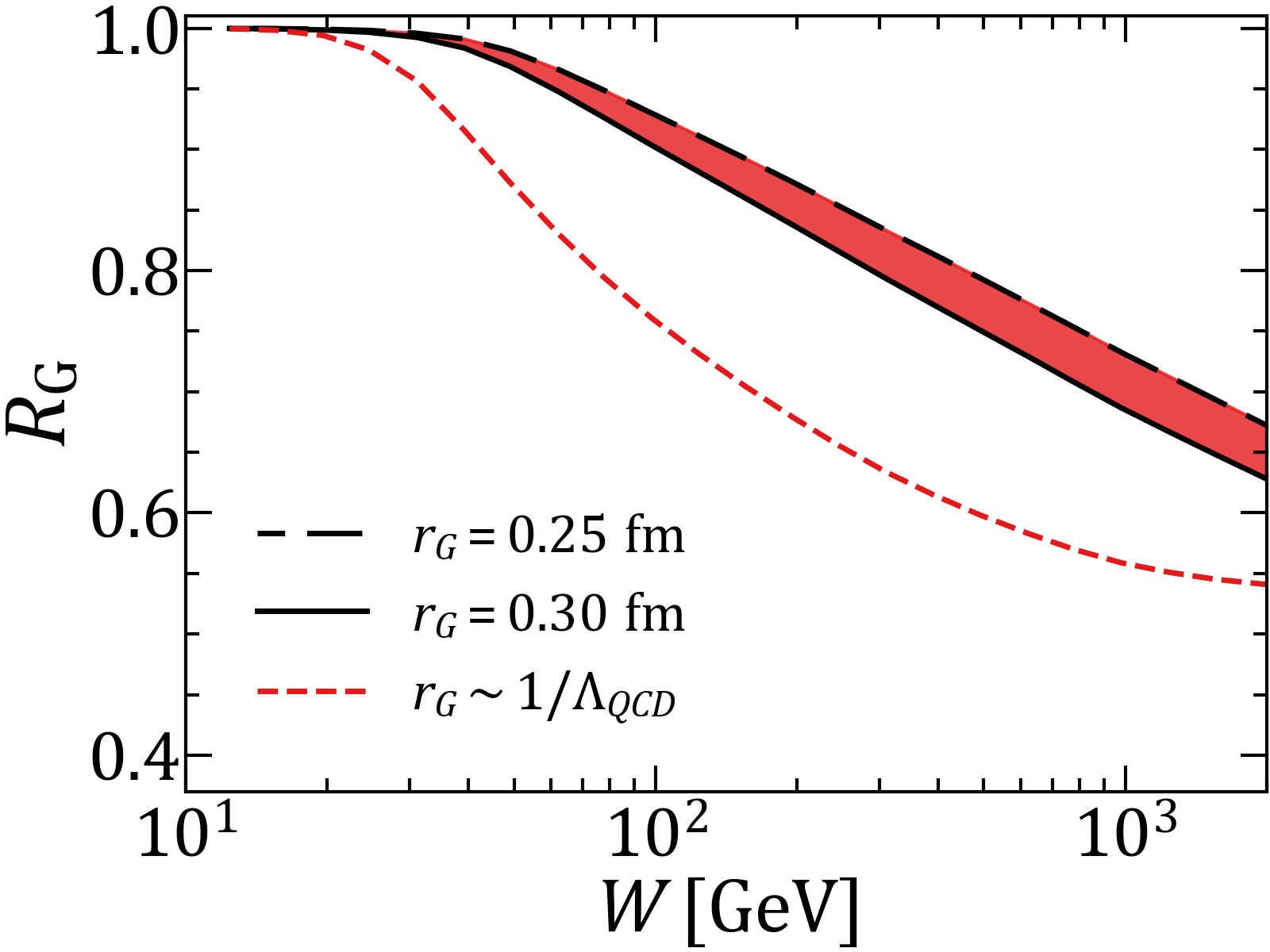}
\\
\Caption{
\label{Fig-RG}
(Color online) 
    Gluon shadowing factor $R_G$ for the coherent photoproduction of $\Jpsi$ off a lead target as a function of the nuclear impact parameter $b$ at several fixed values of the c.m. photon energy $W$ (left), and as a function of $W$ at a photon virtuality $Q^2 = 0$ (right). 
    The upper and lower bounds of the shaded bands correspond to gluon propagation
    radii of $r_G = 0.25\,\fm$ (long-dashed lines) and $r_G = 0.30\,\fm$ (solid lines), respectively.
    The dashed line in the right panel represents our predictions for the $R_G$ factor 
    calculated at an $r_G$ value corresponding to the confinement radius.  
     }
\EF

In the left panel of Fig.~\ref{Fig-RG} we present the nuclear impact parameter $b$
dependence of the $R_G$ factor for the lead target at
several fixed values of $W$ covering the energy range
from experiments at Relativistic Heavy Ion Collider (RHIC) up to values accessible by the planned LHeC.
Here the values of $W=31, 98, 310, 980$, and $3100\,\GeV$ correspond to
$x = 10^{-2}, 10^{-3}, 10^{-4}, 10^{-5}$, and $10^{-6}$, respectively.
The right panel of Fig.~\ref{Fig-RG} shows the $b$-integrated 
gluon shadowing for the lead target at the photon virtuality
$Q^2 = 0$.
The rise in $r_G$ from the value $0.25\,\fm$
to $0.30\,\fm$ corresponds to variation of $R_G$-values from the upper down to lower
bounds of shaded bands, as depicted in Fig.~\ref{Fig-RG}.
One can see that larger values of the gluon propagation radius lead
to a stronger gluon shadowing effect.

As the next step, we test a sensitivity of 
the leading-twist shadowing corrections
to a magnitude of the gluon propagation radius $r_G$, going beyond
the realistic values $\sim 0.25\text{--} 0.30\,\fm$ supported by data.
In the right panel of Fig.~\ref{Fig-RG} we show by the dashed line
the values of the $R_G$ factor at much higher
value of $r_G$ corresponding to the confinement radius $\sim 1/\Lambda_{QCD}$.
One can see that a rise in $r_G$-value  
causes a significant enhancement of leading-twist
corrections, what is demonstrated 
as a difference between the shaded band and the dashed line.

%
%
%
\section{QCD NON-LINEAR EFFECTS: GF FORMALISM VS. BK EQUATION}
\label{res}
%
%
%

For a comprehensive comparison of the strength of non-linear effects between 
the Green's function formalism and the QCD evolution described by the BK equation,
we investigate the coherent photoproduction of charmonia off a lead target. 
Such a comparison is strongly motivated by the
extended kinematic range covered
by future experiments at the LHeC,
reaching large photon energies up to $W=1500\text{--} 2000\,\GeV$.
In this ultra-high 
energy regime, gluon saturation effects are expected to manifest themselves, 
significantly modifying the energy evolution of charmonium production yields 
depending on the actual magnitude of the NL corrections. 
Furthermore, the exclusive 
photoproduction of heavy quarkonia serves as an ideal channel to minimize 
theoretical uncertainties, as demonstrated in
Ref.~\cite{Cepila:2019skb}.

In our analysis of non-linear effects we performed 
calculations of the coherent cross section on lead target
according 
to Eqs.~\eqref{sigcoh}, \eqref{gf-coh} and \eqref{f1}.
These calculations were based on 
the rigorous Green's function formalism described in Sec.~\ref{gf},
where the evolution equation for the Green's function
is governed by Eq.~\eqref{schroedinger}. 
Our choice of a realistic 
POW $c$-$\bar{c}$ 
interaction potential in the rest frame
required a numerical solution of this equation using the algorithm from 
Ref.~\cite{Nemchik:2003wx}.
%
When performing the Lorentz transformation to the LF frame, we fully incorporated 
the quark Melosh spin rotation corrections, as specified in Sec.~\ref{gf}. 
 The exact treatment of quantum coherence 
effects within the GF formalism for the lowest $|Q\bar{Q}\rangle$ Fock state of the 
photon was extended to higher multi-gluon Fock states at large photon energies using the 
same path-integral technique outlined in Sec.~\ref{gs}.

In all numerical calculations, we employed a combination of the realistic POW $c$-$\bar{c}$ 
interaction potential and the recent GBS parametrization for 
$\sigma_{Q\bar{Q}}(r, W)$ introduced in Sec.~\ref{gf}.
This coupling leads to the 
best description of the available experimental data on charmonium photoproduction 
off both proton and lead targets, as demonstrated in 
Figs.~\ref{Fig-psi-p} and \ref{Fig-psi-A}.
For the process $\gamma~p \to \Jpsi~p$, we adopted the slope parameter from 
Ref.~\cite{Cepila:2019skb},
parameterized as 
$B_{\Jpsi}(W) = B_0 + 4 \alpha'(0) \ln(W/W_0)$, 
with parameters $B_0 = 4.4\,\GeV^{-2}$, $W_0 = 90\,\GeV$, and $\alpha'(0) = 0.133\,\GeV^{-2}$ 
obtained from fits to experimental data.
For the photoproduction of $\psip$ we included a correction,
$\Delta(W) = 0.24 \,-\, 0.08 \;\mathrm{ln} \Big(\frac{W}{W_0}\Big)\:\mathrm{GeV}^{-2}$,
as a manifestation of the node effect and modifying slightly the elastic slope $B_{\psip}$
with respect to $B_{\Jpsi}$ via $B_{\psip}(W) = B_{\Jpsi}(W) - \Delta(W)$.

\BF
%
\includegraphics[height=6.5cm]{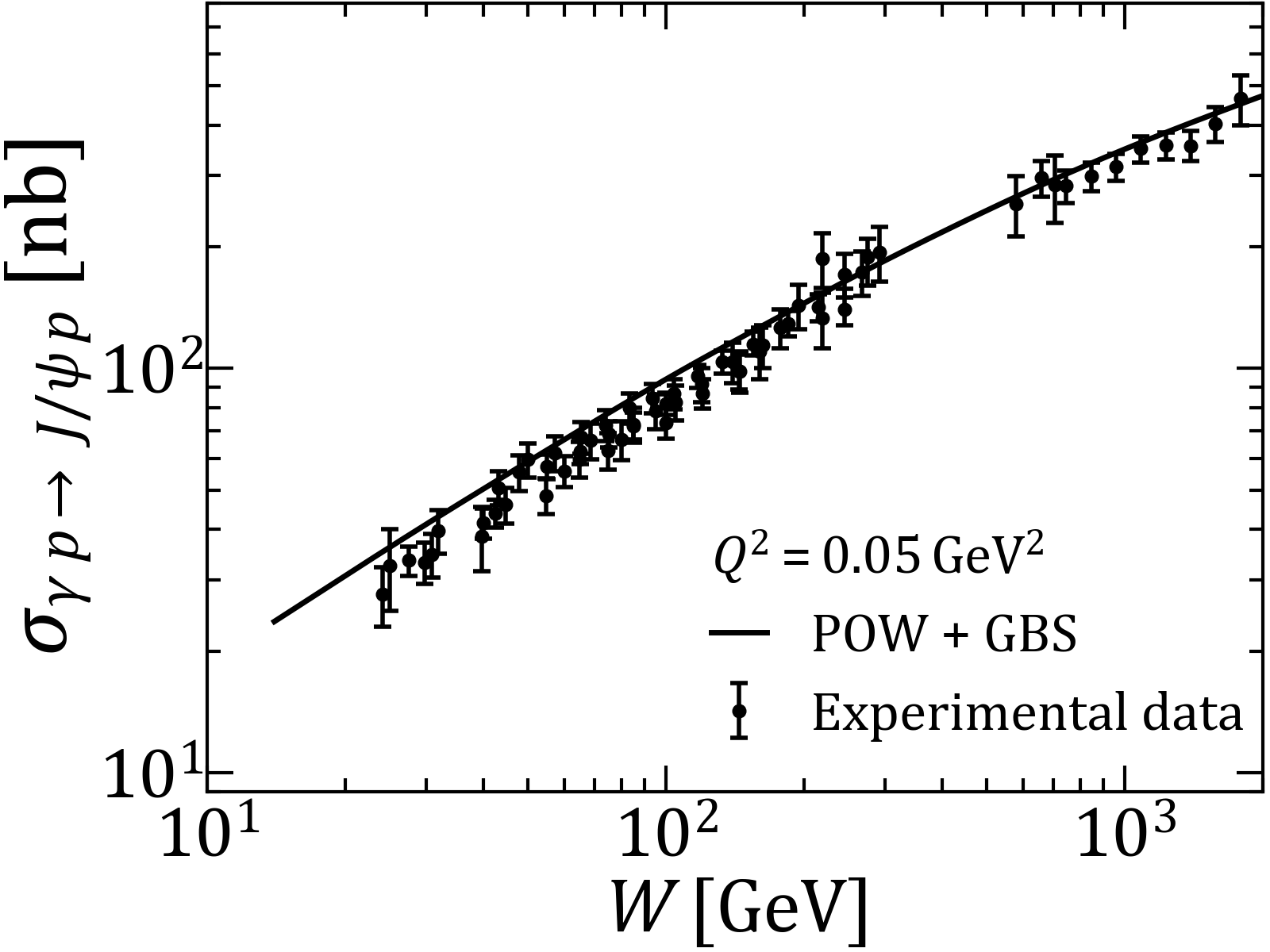}
\hspace*{0.10cm}
\includegraphics[height=6.5cm]{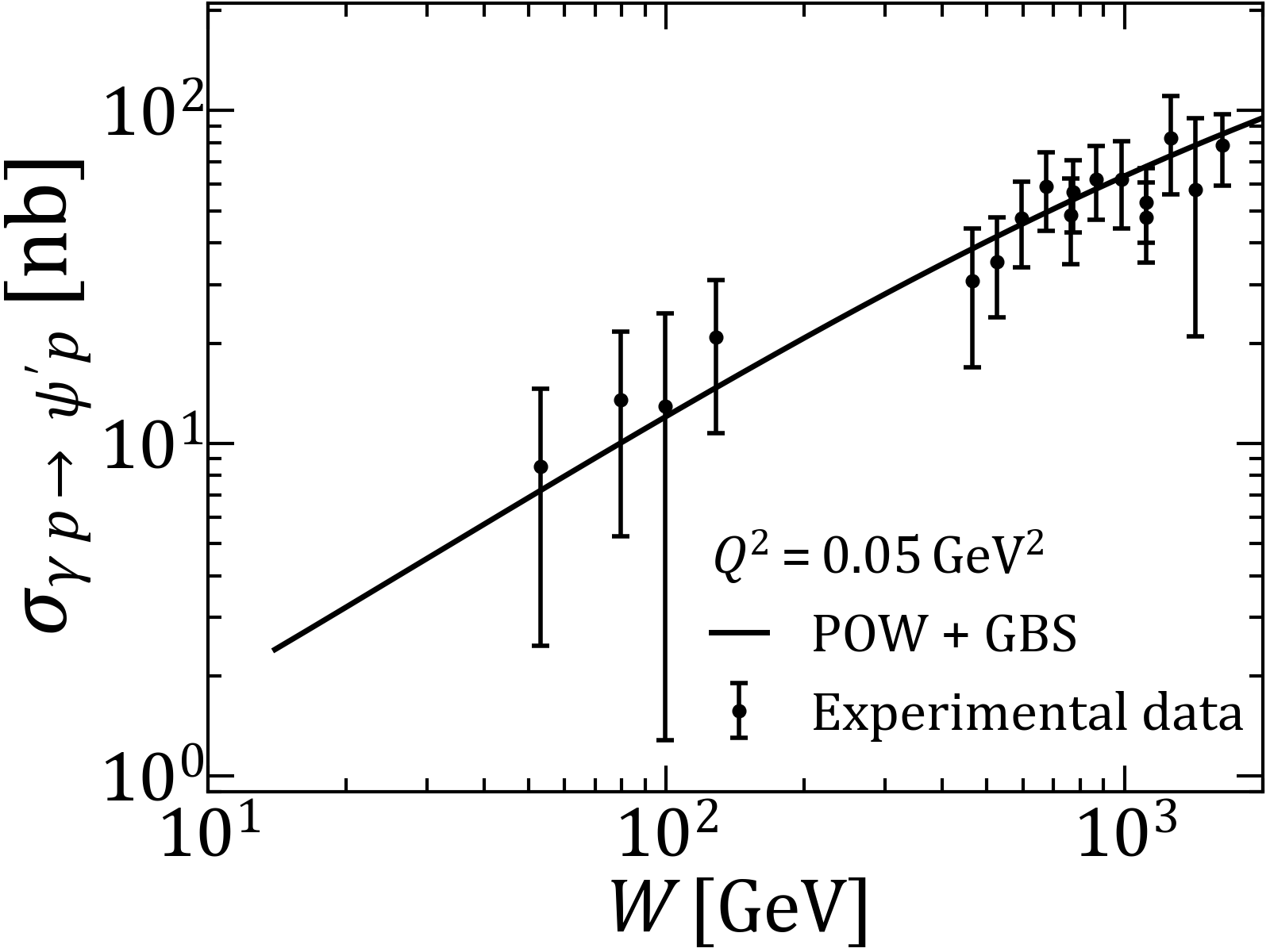}
\\
\Caption{
\label{Fig-psi-p}
(Color online) 
    The model predictions for the photoproduction of $\Jpsi$ (left) and $\psip$ (right)
    off a proton target as a function of the c.m. photon energy $W$.
     The experimental data for $\Jpsi$ are taken from the
     LHCb \cite{LHCb:2018rcm}, ALICE \cite{ALICE:2014eof}, 
     H1 \cite{H1:2005dtp,H1:2013okq}, and  ZEUS \cite{ZEUS:2002wfj} Collaborations.
     The experimental values for $\psip$ were obtained by the
     LHCb\cite{LHCb:2018rcm} and H1\cite{H1:2002yab} Collaborations.      
     }
\EF

\BF
%
\includegraphics[height=6.5cm]{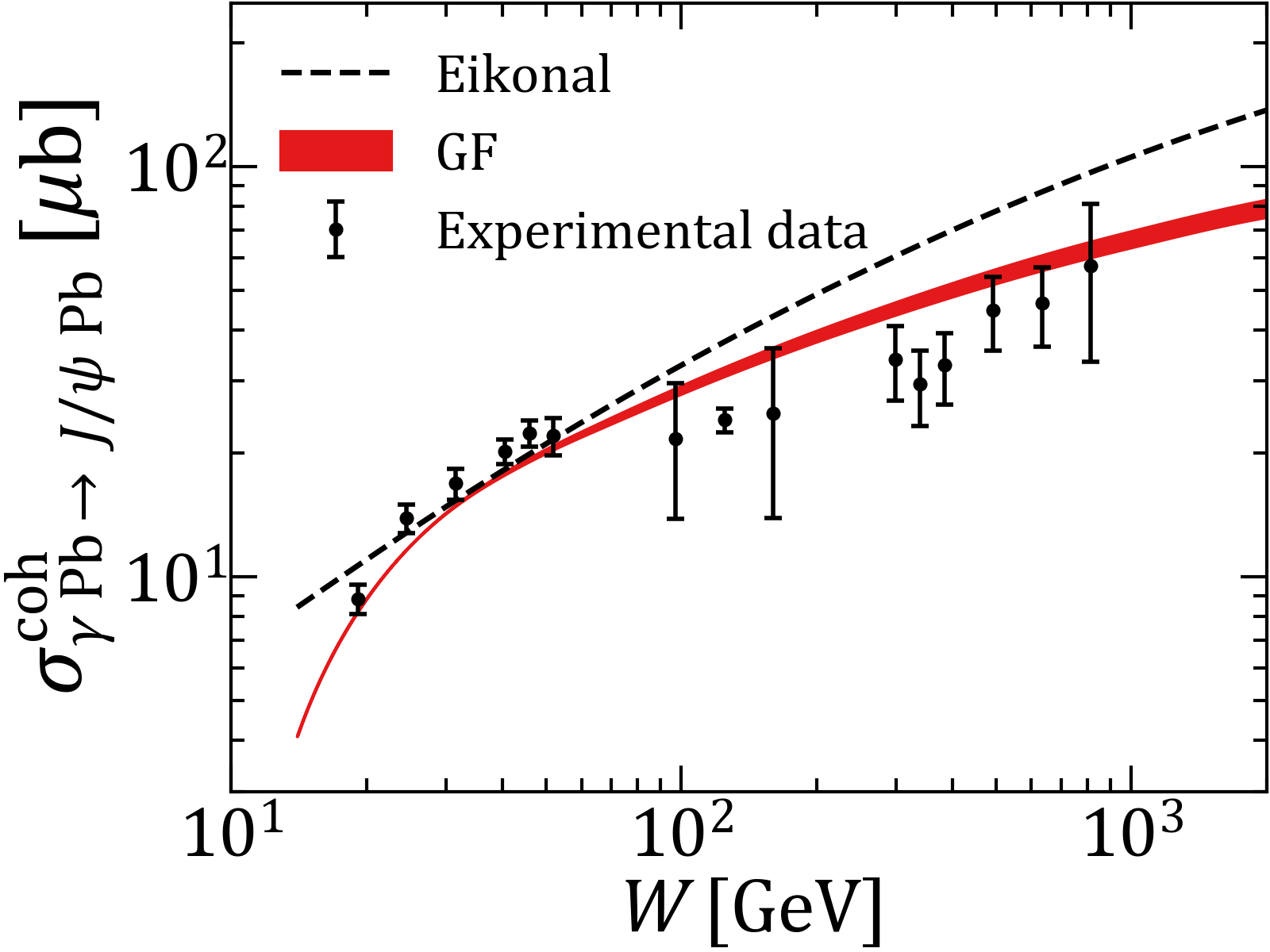}
\hspace*{0.10cm}
\includegraphics[height=6.5cm]{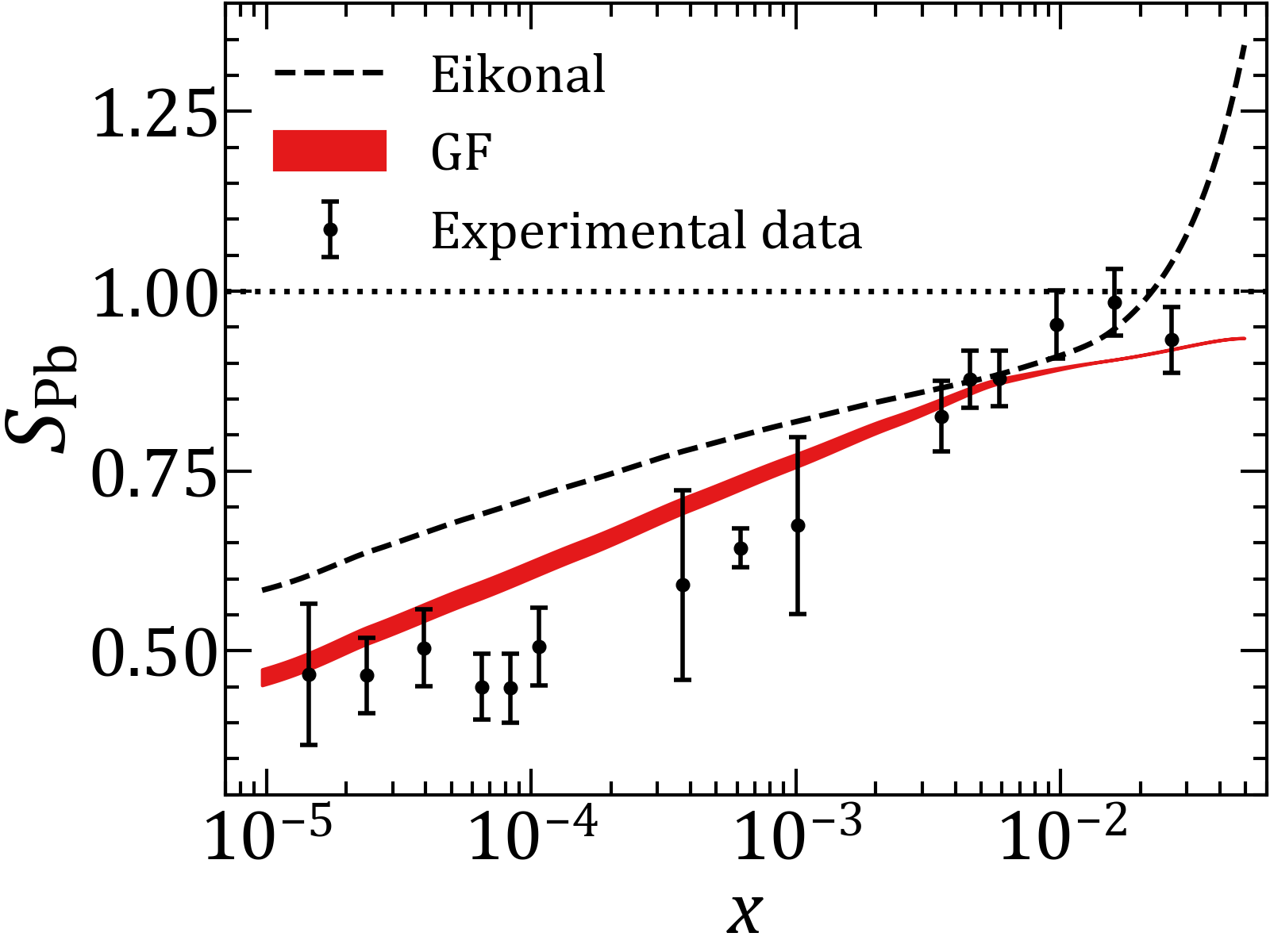}
\\
\Caption{
\label{Fig-psi-A}
(Color online) 
     Total cross section of the coherent photoproduction of $J/\psi$ off a lead target as a function of the c.m. energy $W$  (left) and the nuclear shadowing factor $S_{\mathrm{Pb}}$ as a function of Bjorken-$x$ (right). 
     The dashed lines correspond to calculations in the ``frozen'' approximation
     ($l_c^{Q\bar Q}\gg R_A$)
     given by Eq.~\eqref{coh-lcl}. The shaded bands additionally include the leading-twist
     shadowing corrections at large $W$ and the reduced higher-twist corrections at small $W$. 
     The upper and lower bounds of the shaded bands correspond to gluon propagation radii 
     of $r_G = 0.25\,\fm$ and $r_G = 0.30\,\fm$, respectively.
     The experimental data are taken from the ALICE \cite{ALICE:2023jgu} and CMS \cite{CMS:2023snh} Collaborations.
     }
\EF

Figure~\ref{Fig-psi-A} illustrates how our model predictions based on the Green's
function formalism describe the experimental data on coherent photoproduction of charmonia
off a lead target reported by the ALICE 
\cite{ALICE:2023jgu} 
and CMS 
\cite{CMS:2023snh} 
Collaborations.
While the left panel presents our calculations for the total cross section as a 
function of the c.m.\ photon energy $W$, the right panel exhibits the Bjorken-$x$ 
dependence of the nuclear suppression factor $S_{\text{Pb}}$ proposed in 
Ref.~\cite{Guzey:2013xba} 
for the study of nuclear shadowing. 
This factor is defined as,
%
\BA
    S_{\mathrm{Pb}}(x) 
    = 
    \sqrt{\frac{\sigma^{coh}_{\gamma \,\mathrm{Pb}\rightarrow J/\psi\,\mathrm{Pb}}(x)}{\sigma_{\gamma \,\mathrm{Pb}\rightarrow J/\psi\,\mathrm{Pb}}^{IA}(x)}}\,,
\EA
%
where $\sigma^{coh}_{\gamma \,\mathrm{Pb}\rightarrow J/\psi\,\mathrm{Pb}}(x)$ corresponds to our model predictions and $\sigma^{IA}_{\gamma \,\mathrm{Pb}\rightarrow J/\psi\,\mathrm{Pb}}(x)$
represents the cross section calculated in the impulse approximation  \cite{Guzey:2013xba}.

As evident from Fig.~\ref{Fig-psi-A}, 
the high-energy LCL eikonal approximation, $l_c^{Q\bar{Q}} \gg R_A$
(see also Eq.~\eqref{coh-lcl} ), 
given by dashed lines, fails to provide an adequate description of the data at 
large c.m.\ photon energies $W \gtrsim 100\,\GeV$ (or equivalently, at small 
$x \lsim 3 \times 10^{-3}$). 
The explicit inclusion of the leading-twist gluon shadowing 
corrections provides a good description of the LHC data, 
as shown by the shaded bands.
The upper and lower bounds of these shaded bands correspond to gluon 
propagation radii of $r_G = 0.25\,\fm$ and $r_G = 0.30\,\fm$, respectively.
Furthermore, 
the left panel of Fig.~\ref{Fig-psi-A} clearly demonstrates the impact of reduced quantum coherence effects for the lowest Fock state of the photon toward smaller values of $W$, leading 
to a distinct drop in $\sigma^{\text{coh}}_{\gamma \text{Pb} \to J/\psi \text{Pb}}$ 
relative to the ``frozen'' eikonal approximation.
This allows us to conclude that the higher- and 
leading-twist shadowing corrections naturally embedded within the GF formalism provide 
predictions for charmonium yields that are in good agreement with lead-target 
data across the entire kinematic range $10\,\GeV \lsim W \lsim 2000\,\GeV$. 
This establishes 
a reliable baseline for analyzing the relative significance of the NL effects.

\BF
%
\includegraphics[height=6.5cm]{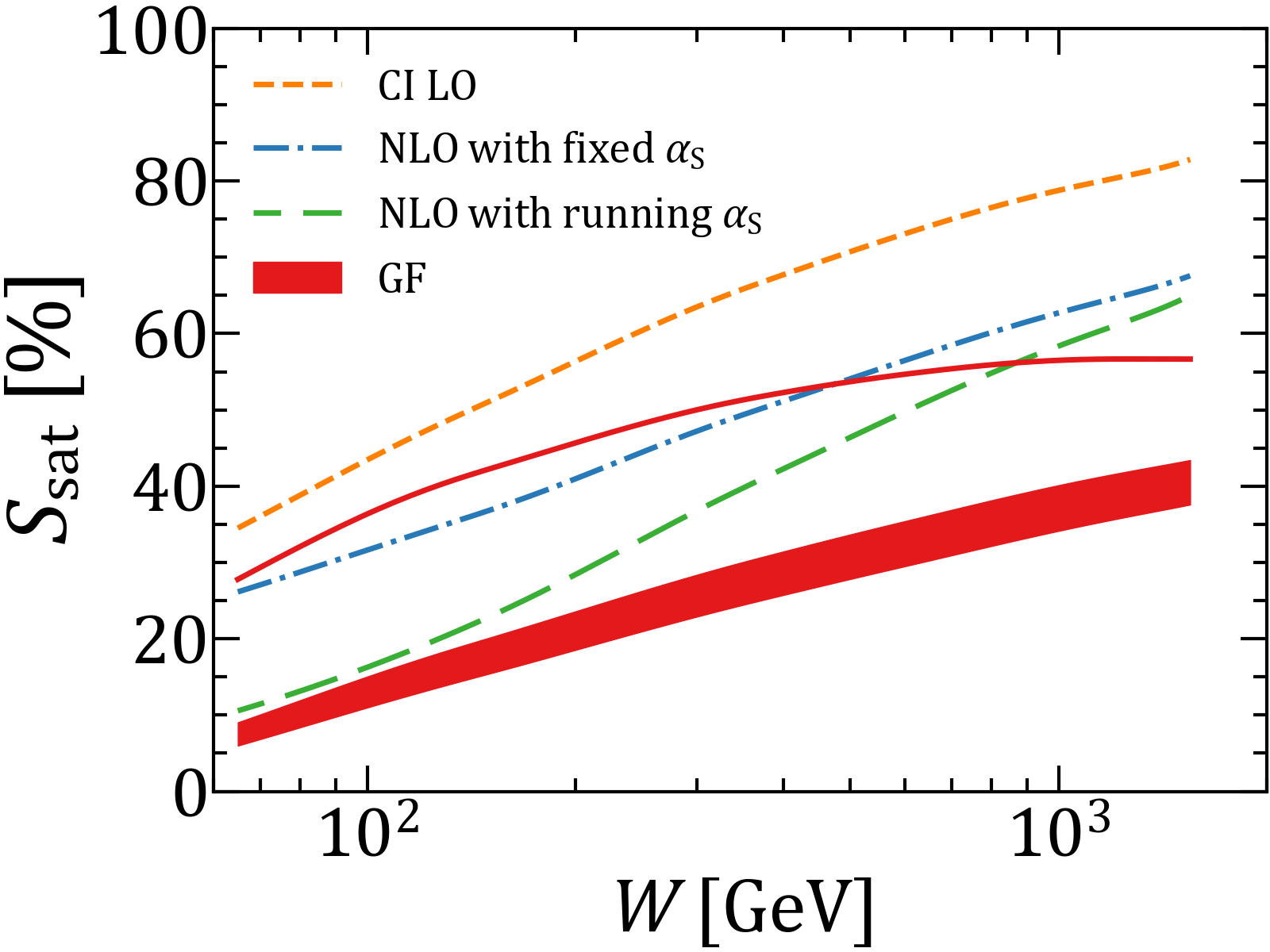}
\hspace*{0.10cm}
\includegraphics[height=6.45cm]{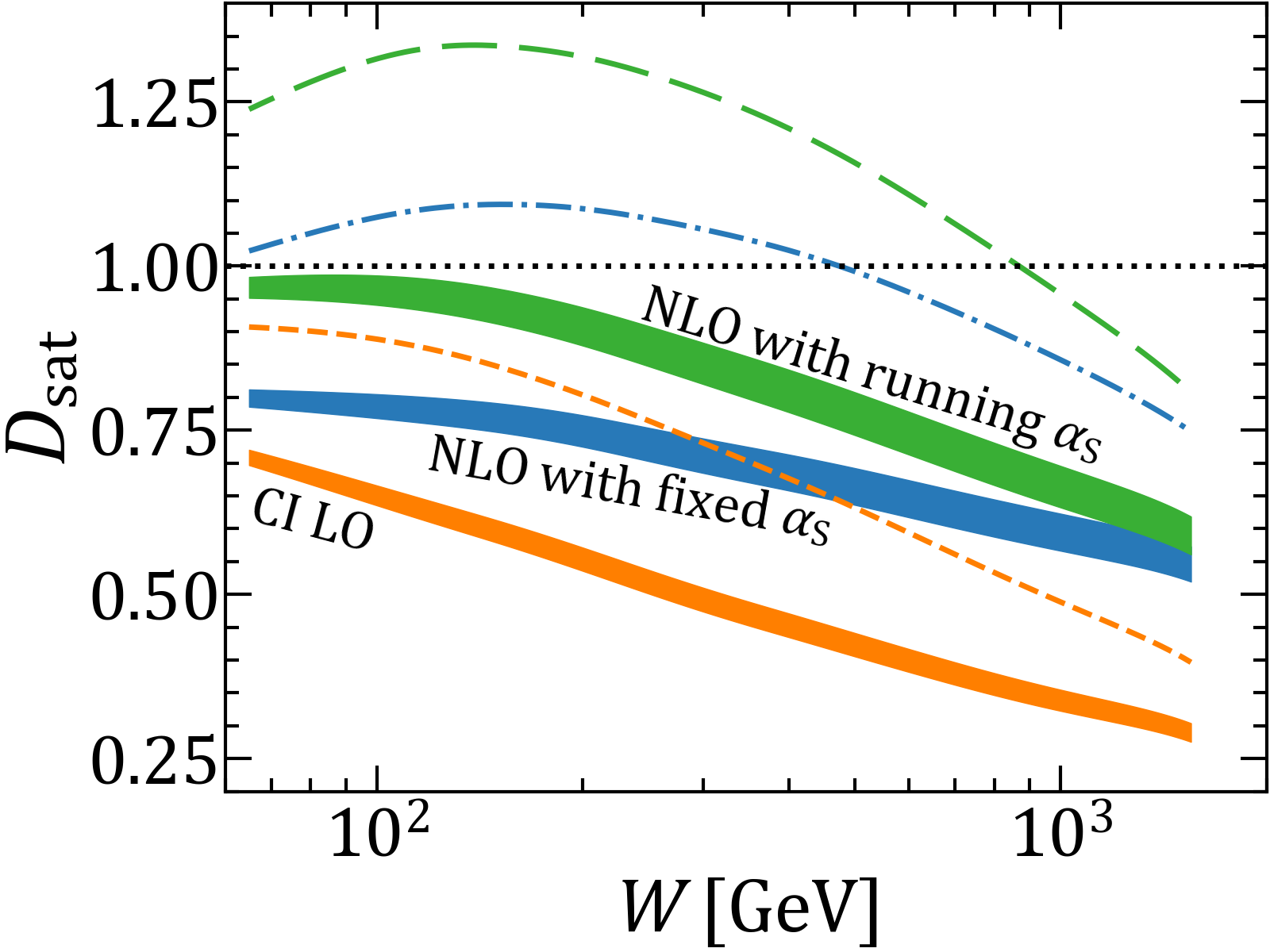}
\\
\Caption{
\label{Fig-NL-effects}
(Color online) 
     Relative impact of non-linear effects on charmonium photoproduction off a lead
     target as a function of the c.m. photon energy $W$ in terms of the variable 
     $\mathcal{S}_{\text{sat}}(W)$ 
     given by Eq.~\eqref{Ssat} (left) and the variable $\mathcal{D}_{\text{sat}}(W)$ 
     defined by Eq.~\eqref{Dsat} (right). 
     In the left panel, the results from the GF formalism (shaded band) are compared with
     values based on the solutions of the LO (dashed line) and the NLO (dot-dashed and long-dashed lines) BK evolution equation from Refs.~\cite{Cepila:2025ujv,Cepila:2024qge}.
     The solid line represents predictions from the GF formalism adopting a large gluon propagation radius $r_G \sim 1/\Lambda_{\text{QCD}}$.
     In the right panel, the 
     shaded bands correspond to
     the ratios 
     $\mathcal{R}_{\text{sat}}^{\text{LO}}/\mathcal{R}_{\text{sat}}^{\text{GF}}$ and 
     $\mathcal{R}_{\text{sat}}^{\text{NLO}}/\mathcal{R}_{\text{sat}}^{\text{GF}}$
     obtained with two variants of the running coupling $\alpha_S$,
     where the ratio $\mathcal{R}_{\text{sat}}$ is defined in Eq.~\eqref{Ssat}.
     The upper and lower bounds of the shaded bands correspond to gluon propagation radii 
     of $r_G = 0.30\,\fm$ and $r_G = 0.25\,\fm$, respectively.
     The long-dashed, dash-dotted, and dashed lines correspond to 
     $\mathcal{D}_{\text{sat}}(W)$ values where the denominator is calculated 
     using a large confinement-scale gluon propagation radius $r_G \sim 1/\Lambda_{\text{QCD}}$.
     }
\EF

In the left panel of Fig.~\ref{Fig-NL-effects}, 
we analyze the relative impact of the QCD non-linear 
effects in the coherent photoproduction of charmonia off a lead target, quantified by 
the variable $\mathcal{S}_{\text{sat}}(W)$,
%
\BE
\label{Ssat}
\mathcal{S}_{\text{sat}}(W) = \bigl| \mathcal{R}_{\text{sat}}(W) - 1 \bigr| = \left| \frac{\sigma_{\text{coh}}^{\text{Pb}}(\text{NL}) - \sigma_{\text{coh}}^{\text{Pb}}(0)}{\sigma_{\text{coh}}^{\text{Pb}}(0)} \right| \times 100\,\% \,,
\EE
%
where $\sigma_{\text{coh}}^{\text{Pb}}(\text{NL})$ and $\sigma_{\text{coh}}^{\text{Pb}}(0)$ 
denote the nuclear cross sections calculated with and without the NL terms, respectively. 
Here, we compare our GF calculations (shaded band) with the results of 
Refs.~\cite{Cepila:2025ujv,Cepila:2024qge}
obtained from the solutions of the BK 
evolution equation at leading order (LO, dashed line) and next-to-leading order (NLO, 
dot-dashed and long-dashed lines).
While 
Ref.~\cite{Cepila:2025ujv} 
explicitly provided 
the LO calculations with and without the NL term in the BK equation, 
the NLO results presented in 
Ref.~\cite{Cepila:2024qge} 
required an indirect extraction of the relative 
NL contribution based on the LO and NLO rapidity evolution of the saturation scale $Q_s(Y)$ from $Y = 0$ to $Y = 10$.

We adopted the reported results from 
Ref.~\cite{Cepila:2024qge} 
for the LO and NLO 
evolution of $Q_s(Y)$ on a proton target, which grows from an initial value of 
$Q_{s0} \approx 0.4\,\GeV$ at $Y = 0$ up to $Q_s^{\text{LO}} \approx 1.05\,\GeV$ and 
$Q_s^{\text{NLO}} \approx 0.70\,\GeV$ at $Y = 10$. 
These values were subsequently rescaled 
for the lead target utilized in our paper.
Following 
Ref.~\cite{Kopeliovich:2010aa},
the nuclear saturation scale $Q_{s,A}$ is related to the nucleon scale $Q_s$ via
%
\BE
Q_{s,A}^2(\vec{b}) = \frac{1}{2} Q_s^2 \sigma_0 T_A(\vec{b}) \,,
\label{eq:QsA-scale}
\EE
%
where $\sigma_0$ represents the saturation level of the dipole cross section.
For our 
evaluation, we use the mean thickness $\langle T_A(\vec{b}) \rangle_b = \int d^2b \, T_A^2(\vec{b})/A$ 
and fix $\sigma_0 = 27.32\,\text{mb}$ from the GBS parametrization. 
This yields 
$Q_{s,\text{Pb}} = 1.46\,Q_s$, corresponding to an initial value of $Q_{s0,\text{Pb}} = 0.58\,\GeV$ 
at $Y = 0$, and evolved values of $Q_{s,\text{Pb}}^{\text{LO}} \approx 1.53\,\GeV$ and 
$Q_{s,\text{Pb}}^{\text{NLO}} \approx 1.02\,\GeV$ at $Y = 10$.

Assuming a constant value for the QCD coupling (cc)
independent of $Q_s^2$, 
the $Y$-evolution of the saturation scale
can be simply parametrized as,
%
\BE
Q^2_{s,Pb}(Y) = Q^2_{so,Pb}\exp[~\lambda\cdot Y~]\,,
\EE
%
allowing us to extract the evolution parameters $\lambda_{LO} = 0.194$ and 
$\lambda_{NLO} = 0.113$.
This leads to a fixed-coupling NLO-to-LO 
evolution rate (speed) ratio of:
%
\BE
\mathcal{R}_{\text{cc}}(Y) = \frac{\left[Q_{s,\text{Pb}}^2(Y)\right]_{\text{NLO}}}{\left[Q_{s,\text{Pb}}^2(Y)\right]_{\text{LO}}} = \exp\bigl[ (\lambda_{\text{NLO}} - \lambda_{\text{LO}})
\cdot Y \bigr] = \exp( -0.081\cdot Y ) \,.
\EE
%
Performing a similar analysis with a running coupling $\alpha_S(Q_s^2)$ (rc), 
the saturation scale behaves as,
%
\BE
Q^2_{s,Pb}(Y) = Q^2_{so,Pb}\exp[~\sqrt{\delta\cdot Y}]~\,,
\EE
%
yielding $\delta_{\text{LO}} = 0.376$ and $\delta_{\text{NLO}} = 0.127$
with a corresponding NLO-to-LO evolution rate (speed) ratio of
%
\BE
\mathcal{R}_{\text{rc}}(Y) = \exp\left[ \bigl(\sqrt{\delta_{\text{NLO}}} - \sqrt{\delta_{\text{LO}}}\bigr)\cdot \sqrt{Y} \right] = \exp( -0.256 \cdot\sqrt{Y} ) \,.
\EE
%
Utilizing these parameterizations, the NLO results can be systematically reconstructed 
from the collinearly improved LO BK calculations, as displayed in
Fig.~\ref{Fig-NL-effects}.
The upper and lower bounds of the shaded
band in the left panel of Fig.~\ref{Fig-NL-effects} correspond 
to gluon propagation radii of 
$r_G = 0.30\,\fm$ and $r_G = 0.25\,\fm$, respectively.
Figure~\ref{Fig-NL-effects} demonstrates that calculations based on the 
standard BK evolution generate a significantly higher relative contribution 
from NL effects compared to GF formalism.
Although the NLO corrections visibly damp the evolution rate and reduce the overall 
magnitude of these effects relative to the LO BK baseline, the discrepancy between 
the BK and GF predictions remains highly significant.

These structural differences can be alternatively scrutinized via the ratio 
$\mathcal{D}_{\text{sat}}(W)$, 
defined as,
%
\BE
\mathcal{D}_{\text{sat}}(W) = \frac{\mathcal{R}_{\text{sat}}^{\text{BK}}(W)}{\mathcal{R}_{\text{sat}}^{\text{GF}}(W)} \,,
\label{Dsat}
\EE
%
where $\mathcal{R}{\text{sat}}(W) = \sigma_{\text{coh}}^{\text{Pb}}(\text{NL}) / \sigma_{\text{coh}}^{\text{Pb}}(0)$.
The much stronger onset of NL effects predicted by both the LO and NLO BK equations
relative to the GF framework results in $\mathcal{D}{\text{sat}}(W) \lesssim 1$, as shown
in the right panel of Fig.~\ref{Fig-NL-effects}.
Moreover, the steady decrease of $\mathcal{D}{\text{sat}}(W)$
with energy confirms that the energy evolution rate of NL effects is inherently stronger
in the BK framework than in the path-integral approach.
The right panel of Fig.~\ref{Fig-NL-effects}
also highlights that NLO corrections temper this evolution rate, as indicated by the
differences between the lower and the middle/upper shaded bands

Additionally, the left panel of Fig.~\ref{Fig-NL-effects} illustrates how the gluon propagation radius regulates the magnitude of non-linear corrections.
Within the GF formalism, increasing
the gluon propagation radius to a confinement-scale value ($r_G \sim 1/\Lambda_{\text{QCD}}$)
leads to substantially larger $\mathcal{S}{\text{sat}}$ values, denoted by the solid line,
thereby narrowing the gap between the GF and BK predictions.
Conversely, the right
panel of Fig.~\ref{Fig-NL-effects} shows that such an increase in $r_G$ can induce a counterintuitive inversion where $\mathcal{D}{\text{sat}} > 1$, 
meaning the onset of NL effects becomes
stronger in the path-integral framework than in the NLO BK results. 
Nevertheless,
these GF-BK discrepancies gradually diminish toward higher photon energies, and at
$W \gtrsim 600\text{--}800\,\GeV$, the standard pattern with $\mathcal{D}_{\text{sat}} < 1$
is fully restored.

These findings allow us to identify two fundamental shortcomings 
of standard applications of the LO and NLO BK evolution equations 
to nuclear targets. 
\\
First, even after implementing 
collinear improvements
(see e.g., Ref.~\cite{Cepila:2025ujv} and references therein),
the BK kernel deals with a weak quark-gluon interaction strength 
associated with an unphysically large gluon propagation radius ($r_G \sim 1/\Lambda_{\text{QCD}}$). 
A subsequent rapid rise of the quark-gluon interaction strength, corresponding to 
a realistic magnitude of $r_G \sim 0.25\text{--}0.30\,\fm$ supported by data on gluon 
radiation
\cite{Kopeliovich:1999am,Kopeliovich:2007pq},
significantly reduces the gluon shadowing (and hence also NL) effects
(see the difference between dashed line and the shaded band in the right panel of Fig.~\ref{Fig-RG}). 
This means that a proper kernel in the BK equation, 
accounting for a quark-gluon interaction whose strength 
is related to a realistic gluon propagation radius, 
is unavoidable for a more correct incorporation of shadowing effects. 
Nevertheless, the sensitivity of both the GF and BK predictions for 
$\sigma_{\text{Pb}}^{\text{coh}}(W)$ to the gluon propagation 
radius slowly decreases with $W$, which is consistent with the 
$1/k$ dependence of the quark-gluon interaction strength 
[see Eq.~\eqref{Re-Vgg}]. 
This leads to a gradual convergence at large $W$ between our predictions for 
$\mathcal{D}_{\text{sat}}(W)$ utilizing GF calculations with a realistic 
$r_G \sim 0.25\text{--}0.30\,\fm$ and those with a confinement scale 
$r_G \sim 1/\Lambda_{\text{QCD}}$, 
manifested in the right panel of Fig.~\ref{Fig-NL-effects} as the closing gap between the explicit lines and the shaded bands.
\\
Second, combining the BK equation with an eikonalized initial condition
for the dipole-nucleus amplitude implicitly assumes that all multi-gluon photon fluctuations
contribute to nuclear shadowing with maximal strength. 
Consequently, the BK equation
severely overestimates nuclear shadowing at large $W$, which can directly bias searches for genuine gluon saturation. 
The BK evolution equation thus requires substantial
modification to properly embed realistic coherence length constraints 
for individual Fock states.
The corresponding derivation of the modified BK equation
will be presented elsewhere.

\BF
%
\includegraphics[height=8.5cm]{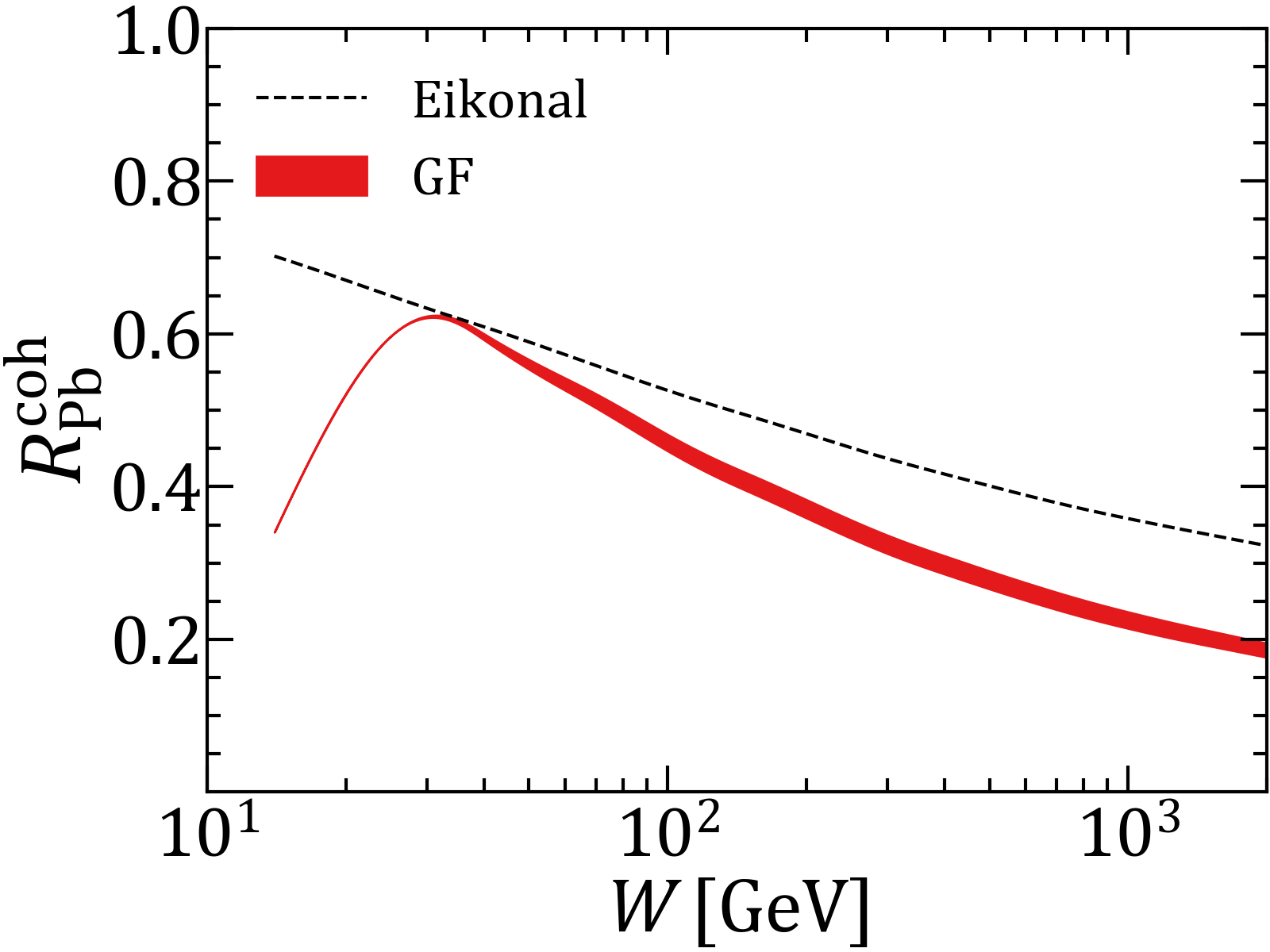}
\\
\Caption{
\label{Fig-R-coh-Pb}
(Color online) 
      Model predictions for the nucleus-to-nucleon cross-section ratio for 
      coherent charmonium photoproduction off a lead target. The dashed line corresponds to the high-energy eikonal approximation, whereas the shaded band represents the GF formalism, which properly incorporates
      quantum coherence effects for both the higher- and leading-twist shadowing corrections.
      The upper and lower bounds of the shaded band correspond to gluon propagation radii 
      of $r_G = 0.25\,\fm$ and $r_G = 0.30\,\fm$, respectively.
     }
\EF

The effective tool for investigation of nuclear effects is 
the nucleus-to-nucleon ratio of cross sections for charmonium 
coherent production, known also as the nuclear modification factor $R_A^{coh}$.
Instead of a standard definition, 
$R_A^{coh} = \sigma_{Pb}^{coh} / A_{Pb}\, \sigma_N$, 
we use the following form of $R_{Pb}^{coh}$ to ensure that the 
nuclear modification factor does not exceed one,
%
\BE
R_{\text{Pb}}^{\text{coh}}(W) 
= 
\frac{\sigma_{\text{coh}}^{\text{Pb}}(W)}
{4\pi B(W) \int d^2b\, T_{Pb}^2(\vec{b})\, \sigma_N(W)}\,,
\EE
%
where $\sigma_N$ is the total charmonium photoproduction cross section on a nucleon target.

In Fig.~\ref{Fig-R-coh-Pb}, we present the prediction for $R_{\text{Pb}}^{\text{coh}}$ as a function of the c.m. energy $W$, 
spanning the kinematic range from RHIC to the planned LHeC. 
The factor $R_{\text{Pb}}^{\text{coh}}$ exhibits a distinct 
non-monotonic energy dependence. 
The NL effects cause a drop in $R_{\text{Pb}}^{\text{coh}}$ toward large
photon energies $W$, whereas a corresponding decrease toward small $W$ 
is driven by the reduced coherence effects 
for the $|Q\bar Q\ra$ Fock state of the photon 
(visible via the differences between
the dashed line and the shaded band in Fig.~\ref{Fig-R-coh-Pb}). 
The significantly stronger onset of NL effects predicted 
by the standard BK equations leads to an additional reduction of
$R_{\text{Pb}}^{\text{coh}}$ at $W = 2000,\GeV$ by a factor of 
two and three for the NLO and LO Kernels, respectively. 
Such a large discrepancy in the factor $R_{\text{Pb}}^{\text{coh}}$
at high photon energies could be successfully probed by future experiments 
at the LHeC.

%
%
%
\section{Conclusions}
\label{conclusions}
%
%
%

In this paper, we have presented a comprehensive comparison of the energy evolution 
of non-linear QCD effects in the coherent photoproduction of charmonia off a 
lead target utilizing two distinct theoretical frameworks. The first method relies 
on calculations based on a rigorous Green's function formalism. Within this 
framework, the corresponding contribution of the leading-twist gluon shadowing
is inherently linked to the NL term in the QCD evolution equations, the explicit 
solutions of which are investigated using the second method. Specifically, we have 
adopted the results for the evolution of the NL term obtained from the numerical 
solutions of the leading-order and next-to-leading-order
Balitsky-Kovchegov equations presented in 
Refs.~\cite{Cepila:2025ujv,Cepila:2024qge}.
Our main observations and conclusions are summarized as follows:

\begin{itemize}

\item 
Considering the lowest $|Q\bar{Q}\ra$ Fock component of the photon 
at the large photon energies accessible by current experiments 
at the LHC and future measurements at the LHeC, 
the coherence length $l_c^{Q\bar{Q}}$ [Eq.~\eqref{lc-qq}] considerably 
exceeds the nuclear radius. 
Consequently, the high-energy LCL eikonal approximation 
[Eq.~\eqref{coh-lcl}], which is naturally reached 
within the GF formalism, 
can be safely adopted to describe nuclear effects, 
leading to the dominance of quark 
shadowing. 
These effects represent a higher-twist contribution to nuclear shadowing 
that is significantly suppressed by the heavy quark mass as $1/m_Q^2$.

\item
The higher multi-gluon Fock components of the photon, 
such as $|Q\bar{Q}G\ra, |Q\bar{Q}GG\ra, \dots$, 
generate the gluon shadowing corrections. 
Compared to the lowest $|Q\bar{Q}\ra$ Fock state, 
these components possess a much shorter coherence length, $l_c^G \ll l_c^{Q\bar{Q}}$, 
and thus require significantly higher energies 
to manifest themselves. 
Since the transverse size of such $Q\bar{Q}\text{--}G$ dipoles 
is nearly scale-independent, the corresponding shadowing correction 
behaves as a leading-twist effect.

\item 
We have found significant differences between the GF formalism 
and the BK evolution predictions for the onset of NL effects, 
which grow systematically with photon energy.
Compared to the GF technique, the relative contribution of 
NL effects to the total coherent charmonium photoproduction 
cross section off a lead target increases by approximately 
$100\%$ and $65\%$ when utilizing results from the LO and 
NLO BK equations, respectively 
(see the left panel of Fig.~\ref{Fig-NL-effects}).

\item 
Analysis of the right panel of Fig.~\ref{Fig-NL-effects} 
demonstrates that the rate of energy 
evolution of the NL effects is substantially higher 
for results based on the explicit 
solutions of the BK equations compared to the GF path-integral formalism.

\item
There are two main shortcomings of the standard BK equation when applied to
nuclear targets. 
First, the collinear-improved BK Kernel 
corresponds to an interaction strength for a quark-gluon system
that is still much weaker than the value related to the gluon
propagation radius $r_G\sim 0.25\text{--}0.30\,\fm$, which is supported 
by data on gluon radiation 
\cite{Kopeliovich:1999am,Kopeliovich:2007pq}.
%
We have found that an increase in the gluon propagation radius leads to
a stronger onset of non-linear effects.
Second, the BK evolution equation, when combined with the initial 
condition for the nuclear dipole amplitude in the high-energy 
eikonal approximation, leads to an overestimation of shadowing effects.
In this framework, the coherence length of each particular Fock state 
of the photon considerably exceeds
the nuclear radius, thereby contributing to shadowing 
with maximal strength.
Consequently, the BK equation requires significant modifications, 
which will be presented in a separate paper.

\item
We have provided predictions for the nuclear modification factor 
$R_{\text{Pb}}^{\text{coh}}$ 
for coherent charmonium photoproduction 
as a function of the center-of-mass photon energy 
$W$, covering the energy range from RHIC to the planned LHeC. 
The reduced coherence effects for the $|Q\bar Q\ra$ Fock state of the photon
cause a suppression toward small photon energies, 
whereas the NL effects generate a significant reduction of $R_{\text{Pb}}^{\text{coh}}$ toward large $W$ (see 
differences between the dashed line and the shaded band in Fig.~\ref{Fig-R-coh-Pb}).
The accelerated onset of NL effects predicted by the solutions of the BK equations leads to an increasing attenuation of $R_{\text{Pb}}^{\text{coh}}$ at $W = 2000\,\GeV$ by an extra factor of two and three for the NLO and LO Kernels, respectively. 
This large discrepancy provides a clear signature that can be easily verified by future experiments at the LHeC.

\item 
Both the sophisticated GF formalism and the BK evolution complement each other, 
representing alternative methods for accounting for non-linear QCD dynamics. 
We expect that they will ultimately yield consistent results;
their future convergence, achieved through an accurate incorporation of 
particular phenomena, such as quantum coherence 
and the realistic interaction strength of a quark-gluon system,
will provide convincing evidence for gluon saturation at high energies.

\end{itemize}

\begin{acknowledgments}

The work of I.K. was partially supported 
by the SGS grant No. SGS26/130/OHK4/2T/14.  
The work of J.N. was partially supported
by the Slovak Funding Agency, Grant No. 2/0020/22.

\end{acknowledgments}




\begin{thebibliography}{99}


\bibitem{Cepila:2025ujv}
J.~Cepila, M.~Matas and M.~Vaculciak,
``Probing nuclear structure with the Balitsky-Kovchegov equation in full impact-parameter dependence,''
Phys. Rev. C \textbf{113}, 055204 (2026).


\bibitem{Cepila:2024qge}
J.~Cepila, J.~G.~Contreras, M.~Matas and M.~Vaculciak,
``Impact-parameter-dependent solutions to the Balitsky-Kovchegov equation at next-to-leading order,''
Phys. Rev. D \textbf{111}, 096015 (2025).


\bibitem{Paakkinen:2026dnh}
P.~Paakkinen and V.~Guzey,
[arXiv:2606.08830 [hep-ph]].


\bibitem{Kopeliovich:1991pu} 
    B.Z.~Kopeliovich and B.G.~Zakharov,
    ``Quantum effects and color transparency in charmonium photoproduction on nuclei,''
    Phys.\ Rev.\ D \textbf{44}, 3466 (1991).

    
\bibitem{Kopeliovich:1998gv}
    B.~Kopeliovich, J.~Raufeisen and A.~Tarasov,
    ``Challenges of nuclear shadowing in DIS,''
    Phys.\ Lett.\ B \textbf{440}, 151-156 (1998)


\bibitem{Kopeliovich:1999am} 
    B.Z.~Kopeliovich, A.~Schafer and A.V.~Tarasov,
    ``Nonperturbative effects in gluon radiation and photoproduction of quark pairs,''
    Phys.\ Rev.\ D \textbf{62}, 054022 (2000).


\bibitem{Kopeliovich:2000ra} 
    B.Z.~Kopeliovich, J.~Raufeisen and A.V.~Tarasov,
    ``Nuclear shadowing and coherence length for longitudinal and transverse photons,''
    Phys.\ Rev.\ C \textbf{62}, 035204 (2000).


\bibitem{Kopeliovich:2001ee}
    B.~Kopeliovich, A.~Tarasov and J.~Hufner,
    ``Coherence phenomena in charmonium production off nuclei at the energies of RHIC and LHC,''
    Nucl. Phys. A \textbf{696}, 669 (2001).


\bibitem{Kopeliovich:2001xj}
    B.~Kopeliovich, J.~Nemchik, A.~Schafer and A.~Tarasov,
    ``Color transparency versus quantum coherence in electroproduction of vector mesons off nuclei,''
    Phys.\ Rev.\ C \textbf{65}, 035201 (2002).


\bibitem{Ivanov:2002kc}
    Y.~Ivanov, B.~Kopeliovich, A.~Tarasov and J.~Hufner,
    ``Electroproduction of charmonia off nuclei,''
    Phys.\ Rev.\ C \textbf{66}, 024903 (2002).


\bibitem{Nemchik:2003wx}
    J.~Nemchik,
     ``Nuclear shadowing in DIS: Numerical solution of evolution equation for the green's function,''
    Phys.\ Rev.\ C \textbf{68}, 035206 (2003).


\bibitem{Kopeliovich:2008ek}
    B.~Kopeliovich, J.~Nemchik, I.~Potashnikova and I.~Schmidt,
    ``Gluon Shadowing in DIS off Nuclei,''
    J.\ Phys.\ G \textbf{35}, 115010 (2008).


\bibitem{Krelina:2020ipn}
    M.~Krelina and J.~Nemchik,
    ``Nuclear shadowing in DIS at electron-ion colliders,''
    Eur. Phys. J. Plus \textbf{135}, 444 (2020).
    

\bibitem{Kopeliovich:2022jwe}
    B.~Z.~Kopeliovich, M.~Krelina, J.~Nemchik and I.~K.~Potashnikova,
    ``Coherent photoproduction of heavy quarkonia on nuclei,''
    Phys. Rev. D \textbf{105}, 054023 (2022).

    
\bibitem{Kopeliovich:2020has}
    B.~Z.~Kopeliovich, M.~Krelina, J.~Nemchik and I.~K.~Potashnikova,
    ``Ultraperipheral nuclear collisions as a source of heavy quarkonia,''
    Phys. Rev. D \textbf{107}, 054005 (2023).


\bibitem{Arneodo:1992wf} 
  M.~Arneodo,
  Phys.\ Rept.\ \textbf{240}, 301 (1994).


\bibitem{Kancheli:1973vc}
   O.~V.~Kancheli,
   ``Inelastic interactions of fast hadrons with nuclei,''
   Pisma Zh. Eksp. Teor. Fiz. \textbf{18}, 465 (1973);
   Sov. Phys. JETP Lett. \textbf{18}, 274 (1973).


\bibitem{Gribov:1983ivg}
   L.~V.~Gribov, E.~M.~Levin and M.~G.~Ryskin,
   ``Semihard Processes in QCD,''
   Phys. Rept. \textbf{100}, 1 (1983).


\bibitem{Mueller:1985wy}
   A.~H.~Mueller and J.~w.~Qiu,
   ``Gluon Recombination and Shadowing at Small Values of x,''
   Nucl. Phys. B \textbf{268}, 427 (1986).


\bibitem{Qiu:1986wh}
   J.~w.~Qiu,
   ``Nuclear Shadowing at Small Values of x,''
   Nucl. Phys. B \textbf{291}, 746 (1987).


\bibitem{Bauer:1977iq}
   T.~H.~Bauer, R.~D.~Spital, D.~R.~Yennie and F.~M.~Pipkin,
   ``The Hadronic Properties of the Photon in High-Energy Interactions,''
   Rev. Mod. Phys. \textbf{50}, 261 (1978);
   [erratum: Rev. Mod. Phys. \textbf{51}, 407 (1979)].


\bibitem{Frankfurt:1988nt}
   L.~L.~Frankfurt and M.~I.~Strikman,
   ``Hard Nuclear Processes and Microscopic Nuclear Structure,''
   Phys. Rept. \textbf{160}, 235 (1988).


\bibitem{Brodsky:1989qz}
   S.~J.~Brodsky and H.~J.~Lu,
   ``Shadowing and Antishadowing of Nuclear Structure Functions,''
   Phys. Rev. Lett. \textbf{64}, 1342 (1990).


\bibitem{Brodsky:2004qa}
   S.~J.~Brodsky, I.~Schmidt and J.~J.~Yang,
   ``Nuclear antishadowing in neutrino deep inelastic scattering,''
   Phys. Rev. D \textbf{70}, 116003 (2004).


\bibitem{Nikolaev:1990ja}
   N.~N.~Nikolaev and B.~G.~Zakharov,
   ``Color transparency and scaling properties of nuclear shadowing in deep inelastic scattering,''
   Z. Phys. C \textbf{49}, 607 (1991).


\bibitem{Melnitchouk:1993vc}
   W.~Melnitchouk and A.~W.~Thomas,
   ``Nuclear shadowing at small x and Q**2,''
   Phys. Lett. B \textbf{317}, 437 (1993).


\bibitem{Nikolaev:1994de}
   N.~N.~Nikolaev, G.~Piller and B.~G.~Zakharov,
   ``Quantum coherence in heavy flavor production on nuclei,''
   J. Exp. Theor. Phys. \textbf{81}, 851 (1995).


\bibitem{Piller:1995kh}
   G.~Piller, W.~Ratzka and W.~Weise,
   ``Phenomenology of nuclear shadowing in deep inelastic scattering,''
   Z. Phys. A \textbf{352}, 427 (1995).


\bibitem{Kopeliovich:1995yr}
   B.~Kopeliovich and B.~Povh,
   ``Scaling variable for nuclear shadowing in deep inelastic scattering,''
   Phys. Lett. B \textbf{367}, 329 (1996).


\bibitem{Kopeliovich:1995ju}
   B.~Kopeliovich and B.~Povh,
   ``Interplay of soft and hard interactions in nuclear shadowing at high q**2 and low x,''
   Z. Phys. A \textbf{356}, 467 (1997).


\bibitem{Piller:1999wx}
   G.~Piller and W.~Weise,
   ``Nuclear deep inelastic lepton scattering and coherence phenomena,''
   Phys. Rept. \textbf{330}, 1 (2000).


\bibitem{LHeC:2020van}
    P.~Agostini \textit{et al.} [LHeC and FCC-he Study Group],
    ``The Large Hadron\textendash{}Electron Collider at the HL-LHC,''
    J. Phys. G \textbf{48}, 110501 (2021).


\bibitem{Nemchik:2025myg}
   J.~Nemchik and J.~{\'O}bertov{\'a},
   ``Searching for gluon saturation effects in the momentum transfer dependence of coherent charmonium electroproduction off nuclei,''
   Phys. Rev. D \textbf{112}, 094005 (2025).


\bibitem{Kopeliovich:2007pq}
   B.~Z.~Kopeliovich, I.~K.~Potashnikova, B.~Povh and I.~Schmidt,
   ``Evidences for two scales in hadrons,''
   Phys. Rev. D \textbf{76}, 094020 (2007).


\bibitem{Gribov:1968jf}
   V.~N.~Gribov,
   ``Glauber Corrections and the Interaction between High-energy Hadrons and Nuclei,''
   Sov. Phys. JETP \textbf{29}, 483 (1969).


\bibitem{Nemchik:2002ug}
    J.~Nemchik,
    ``Incoherent production of charmonia off nuclei as a good tool for study of color transparency,''
    Phys.\ Rev.\ C \textbf{66}, 045204 (2002).


\bibitem{Nemchik:2024lny}
   J.~Nemchik and J.~{\'O}bertov{\'a},
   ``Coherent photo- and electroproduction of charmonia on nuclear targets revisited: Green function formalism,''
   Phys. Rev. D \textbf{110}, 054015 (2024).


\bibitem{Kopeliovich:2001hf}
   B.~Z.~Kopeliovich, J.~Raufeisen, A.~V.~Tarasov and M.~B.~Johnson,
   ``Nuclear effects in the Drell-Yan process at very high-energies,''
   Phys. Rev. C \textbf{67}, 014903 (2003).


\bibitem{Gribov:1981ac}
   L.V. Gribov, E.M.~Levin and M.G.~Ryskin,
   ``Singlet Structure Function at Small x: Unitarization of Gluon Ladders,''
   Nucl. Phys. B \textbf{188}, 555 (1981).


\bibitem{balitsky} 
   I. Balitsky, 
   ``Operator expansion for high-energy scattering,''
   Nucl. Phys. B \textbf{463}, 99  (1996).


\bibitem{kovchegov}
   Y. V. Kovchegov, 
   ``Small-x $F_2$ structure function of a nucleus including multiple pomeron exchanges,''
   Phys. Rev. D \textbf{60}, 034008  (1999).


\bibitem{Mantysaari:2022kdm}
    H.~M\"antysaari and J.~Penttala,
    ``Complete calculation of exclusive heavy vector meson production at next-to-leading order in the dipole picture,''
    JHEP \textbf{08}, 247 (2022).


\bibitem{Eskola:2022vpi}
   K.~J.~Eskola, C.~A.~Flett, V.~Guzey, T.~L{\"o}yt{\"a}inen and H.~Paukkunen,
   ``Exclusive J/{\ensuremath{\psi}} photoproduction in ultraperipheral Pb+Pb collisions at the CERN Large Hadron Collider calculated at next-to-leading order perturbative QCD,''
   Phys. Rev. C \textbf{106}, 035202 (2022).



\bibitem{Eskola:2022vaf}
    K.~J.~Eskola, C.~A.~Flett, V.~Guzey, T.~L\"oyt\"ainen and H.~Paukkunen,
    ``Next-to-leading order perturbative QCD predictions for exclusive J/\ensuremath{\psi} photoproduction in oxygen-oxygen and lead-lead collisions at energies available at the CERN Large Hadron Collider,''
    Phys. Rev. C \textbf{107}, 044912 (2023).


\bibitem{Cepila:2019skb}
    J.~Cepila, J.~Nemchik, M.~Krelina and R.~Pasechnik,
    ``Theoretical uncertainties in exclusive electroproduction of S-wave heavy quarkonia,''
    Eur.\ Phys.\ J.\ C \textbf{79}, 495 (2019).


\bibitem{Krelina:2018hmt}
    M.~Krelina, J.~Nemchik, R.~Pasechnik and J.~Cepila,
    ``Spin rotation effects in diffractive electroproduction of heavy quarkonia,''
    Eur.\ Phys.\ J.\ C \textbf{79}, 154 (2019).


\bibitem{Hufner:2000jb} 
    J.~Hufner, Y.P.~Ivanov, B.Z.~Kopeliovich and A.V.~Tarasov;
    ``Photoproduction of charmonia and total charmonium proton cross-sections,''
    Phys.\ Rev.\ D \textbf{62}, 094022 (2000).


\bibitem{DeJager:1987qc}
    H.De Vries, C.W.De Jager and C.De Vries,
    ``Nuclear charge and magnetization density distribution parameters from elastic electron scattering,''
    Atom.\ Data Nucl.\ Data Tabl.\ \textbf{36}, 495 (1987).


 \bibitem{Martin:1980jx}
    A.~Martin;
    ``A FIT of Upsilon and Charmonium Spectra,''
    Phys.\ Lett.\ B\textbf{93}, 338 (1980).


\bibitem{Barik:1980ai}
    N.~Barik and S.N.~Jena;
    ``Fine - Hyperfine Splittings Of Quarkonium Levels In An Effective Power Law Potential,''
    Phys.\ Lett.\ B \textbf{97}, 265 (1980). 


\bibitem{Terentev:1976jk} 
    M.V.~Terentev;
    ``On the Structure of Wave Functions of Mesons as Bound States of Relativistic Quarks,''
    Sov.\ J.\ Nucl.\ Phys.\ \textbf{24}, 106 (1976)
    [Yad.\ Fiz.\ \textbf{24}, 207 (1976)].    


\bibitem{Krelina:2019egg}
   M.~Krelina, J.~Nemchik and R.~Pasechnik,
    ``$D$-wave effects in diffractive electroproduction of heavy quarkonia from the photon-like $V\rightarrow Q\bar Q$ transition,''
    Eur.\ Phys.\ J.\ C \textbf{80}, 92 (2020).


\bibitem{Kopeliovich:2015qna} 
    B.Z.~Kopeliovich, E.~Levin, I.~Schmidt and M.~Siddikov;
    ``Lorentz-boosted description of a heavy quarkonium,''
    Phys.\ Rev.\ D \textbf{92}, 034023 (2015).


\bibitem{Lappi:2020ufv}
    T.~Lappi, H.~M\"antysaari and J.~Penttala,
    ``Relativistic corrections to the vector meson light FRONT wave function,''
    Phys. Rev. D \textbf{102}, 054020 (2020).


\bibitem{Melosh:1974cu} 
    H.J.~Melosh;
    ``Quarks: Currents and constituents,''
    Phys.\ Rev.\ D \textbf{9}, 1095 (1974).


\bibitem{Golec-Biernat:2017lfv}
    K.~Golec-Biernat and S.~Sapeta,
    ``Saturation model of DIS : an update,''
    JHEP \textbf{03}, 102 (2018).


\bibitem{Nemchik:1996cw}
    J.~Nemchik, N.~N.~Nikolaev, E.~Predazzi and B.~Zakharov,
    ``Color dipole phenomenology of diffractive electroproduction of light vector mesons at HERA,''
    Z.\ Phys.\ C \textbf{75}, 71 (1997).


\bibitem{Bronzan:1974jh}
   J.~B.~Bronzan, G.~L.~Kane and U.~P.~Sukhatme,
   ``Obtaining Real Parts of Scattering Amplitudes Directly from Cross-Section Data Using Derivative Analyticity Relations,''
   Phys. Lett. B \textbf{49}, 272 (1974).


\bibitem{Forshaw:2003ki}
   J.~R.~Forshaw, R.~Sandapen and G.~Shaw,
   ``Color dipoles and rho, phi electroproduction,''
   Phys. Rev. D \textbf{69}, 094013 (2004).

   
\bibitem{Shuvaev:1999ce}
    A.~G.~Shuvaev, K.~J.~Golec-Biernat, A.~D.~Martin and M.~G.~Ryskin,
    ``Off diagonal distributions fixed by diagonal partons at small x and xi,''
    Phys. Rev. D \textbf{60}, 014015 (1999).


\bibitem{Ivanov:2007ms}
   Y.~Ivanov, B.~Kopeliovich and I.~Schmidt,
   ``Vector meson production in ultra-peripheral collisions at LHC,''
   arXiv:{0706.1532 [hep-ph]}.


\bibitem{LHCb:2018rcm}
   R.~Aaij \textit{et al.} [LHCb],
   ``Central exclusive production of $J/\psi$ and $\psi(2S)$ mesons in $pp$ collisions at $\sqrt{s}=13~$TeV,''
   JHEP \textbf{10}, 167 (2018).


\bibitem{ALICE:2014eof}
   B.~B.~Abelev \textit{et al.} [ALICE],
   ``Exclusive $\mathrm{J/}\psi$ photoproduction off protons in ultra-peripheral p-Pb collisions at $\sqrt{s_{\rm NN}}=5.02$ TeV,''
   Phys. Rev. Lett. \textbf{113}, 232504 (2014).


\bibitem{H1:2005dtp}
   A.~Aktas \textit{et al.} [H1],
   ``Elastic J/psi production at HERA,''
   Eur. Phys. J. C \textbf{46}, 585 (2006).


\bibitem{H1:2013okq}
   C.~Alexa \textit{et al.} [H1],
   ``Elastic and Proton-Dissociative Photoproduction of J/psi Mesons at HERA,''
   Eur. Phys. J. C \textbf{73}, 2466 (2013).


\bibitem{ZEUS:2002wfj}
   S.~Chekanov \textit{et al.} [ZEUS],
   ``Exclusive photoproduction of J / psi mesons at HERA,''
   Eur. Phys. J. C \textbf{24}, 345 (2002).


\bibitem{H1:2002yab}
   C.~Adloff \textit{et al.} [H1],
   ``Diffractive photoproduction of psi(2S) mesons at HERA,''
   Phys. Lett. B \textbf{541}, 251 (2002).


\bibitem{ALICE:2023jgu}
   S.~Acharya \textit{et al.} [ALICE],
   ``Energy dependence of coherent photonuclear production of J/{\ensuremath{\psi}} mesons in ultra-peripheral Pb-Pb collisions at $ \sqrt{{\textrm{s}}_{\textrm{NN}}} $ = 5.02 TeV,''
   JHEP \textbf{10}, 119 (2023).


\bibitem{CMS:2023snh}
   A.~Tumasyan \textit{et al.} [CMS],
   ``Probing Small Bjorken-x Nuclear Gluonic Structure via Coherent J/{\ensuremath{\psi}} Photoproduction in Ultraperipheral Pb-Pb Collisions at sNN=5.02{\,}{\,}TeV,''
   Phys. Rev. Lett. \textbf{131}, 262301 (2023).


\bibitem{Guzey:2013xba}
   V.~Guzey, E.~Kryshen, M.~Strikman and M.~Zhalov,
   ``Evidence for nuclear gluon shadowing from the ALICE measurements of PbPb ultraperipheral exclusive $J/{\psi}$ production,''
   Phys. Lett. B \textbf{726}, 290 (2013).


\bibitem{Kopeliovich:2010aa}
   B.~Z.~Kopeliovich, I.~K.~Potashnikova and I.~Schmidt,
   ``Measuring the saturation scale in nuclei,''
   Phys. Rev. C \textbf{81}, 035204 (2010).


\end{thebibliography}
\end{document}